\documentclass[conference,compsoc]{IEEEtran}
\usepackage{booktabs}    
\usepackage{array}       
\usepackage{tabularx}
\usepackage{enumitem}
\usepackage{longtable}
\usepackage{tcolorbox}
\usepackage{xcolor}
\usepackage{adjustbox}
\usepackage{xspace}
\usepackage{mdframed}
\usepackage{minted}  
\usepackage{tcolorbox} 
\tcbuselibrary{skins,breakable,minted}
\usepackage{url}

\usepackage{amsmath}
\usepackage{amssymb}
\newcommand{\codeqlV}{v2.21.2\xspace}

\newcommand{\toolname}{SemTaint\xspace}
\newcommand{\sourceSinkAgentName}{Source/Sink\xspace}
\newcommand{\callgraphAgentName}{CallGraph\xspace}
\newcommand{\flowSummaryAgentName}{Flow Summary\xspace}
\newcommand{\thirdPartyEdgeName}{candidate\xspace}
\newcommand{\ThirdPartyEdgeName}{Candidate\xspace}
\definecolor{VerifyLightRed}{RGB}{255, 104, 103}

\usepackage{listings}

\newcommand{\circletext}[2]{%
  \tikz[baseline=(char.base)]{
    \node[shape=circle, draw=#1, fill=#1, text=white, inner sep=0.5pt] (char) {\footnotesize\bfseries #2};
  }%
}

\newcommand{\squaretext}[2]{%
  \tikz[baseline=(char.base)]{
    \node[shape=rectangle, rounded corners=1.2pt, draw=#1, fill=#1, text=white, inner sep=2pt] (char) {\footnotesize\bfseries #2};
  }%
}

\newcommand{\myparagraph}[1]{\noindent\textbf{#1}.}

\newcommand{\mytitle}[1]{\vspace{5pt}\noindent\textbf{#1}.}
\newcommand{\mysubtitle}[1]{\textbf{#1}.}

\newtcblisting{codeblock}[3][]{%
  listing engine=minted,
  listing only,              
  minted language=#2,
  minted options = {
    breaklines,
    autogobble,
    linenos,
    numbersep=8pt,
    fontsize=\small,
    tabsize=2,
    #1
  },
  colback=gray!2,
  colframe=gray!60,
  coltitle=black,
  title={#3},
  left=8pt, right=8pt, top=6pt, bottom=6pt,
  boxrule=0.6pt, arc=2mm,
  enhanced, breakable
}

\newtcolorbox{keyfindingcolorbox}{
  colback=gray!3,
  colframe=gray!200,
  boxrule=1pt,
  arc=3pt,
  left=2pt,right=2pt,top=2pt,bottom=2pt
}

\newcommand{\valTotal}{10\xspace}

\newcommand{\testingTotal}{162\xspace}
\newcommand{\testingTP}{106\xspace}

\newcommand{\testingRecall}{65.43\%\xspace}

\newcommand{\totalOss}{10\xspace}
\newcommand{\ossResults}{4\xspace}
\newcommand{\ossResultsWord}{four\xspace}

\definecolor{custompurple}{HTML}{7F00FF}
\definecolor{customred}{HTML}{CC0001}
\definecolor{customgreen}{HTML}{038A00}
\definecolor{customorange}{HTML}{FE8701}
\definecolor{custombluedark}{HTML}{014A94}
\definecolor{custombluelight}{HTML}{007FFF}
\definecolor{customblack}{HTML}{000000}

\definecolor{customlightpurple}{HTML}{9933FF}
\definecolor{customlightblue}{HTML}{00CCCC}
\definecolor{custompink}{HTML}{FF33FF}
\definecolor{customdarkred}{HTML}{990000}

\usepackage{tikz}
\usepackage[utf8]{inputenc}
\usepackage{textgreek}
\usepackage{algorithm}
\usepackage{algpseudocode}
\usepackage{textcomp}
\usepackage{caption}
\usepackage{xcolor}
\usepackage{mathtools}
\usepackage{bm}

\usetikzlibrary{shapes.geometric, arrows, positioning, calc, arrows.meta}

\begin{document}

\date{}


\title{\Large \bf Multi-Agent Taint Specification Extraction for Vulnerability Detection}


\author{
\IEEEauthorblockN{Jonah Ghebremichael\IEEEauthorrefmark{1}, Saastha Vasan\IEEEauthorrefmark{2}, Saad Ullah\IEEEauthorrefmark{3}, Greg Tystahl\IEEEauthorrefmark{1}, \\ David Adei\IEEEauthorrefmark{1},  Christopher Kruegel\IEEEauthorrefmark{2}, Giovanni Vigna\IEEEauthorrefmark{2}, William Enck\IEEEauthorrefmark{1}, Alexandros Kapravelos\IEEEauthorrefmark{1}}
\IEEEauthorblockA{\IEEEauthorrefmark{1}North Carolina State University,
\IEEEauthorrefmark{2}University of California, Santa Barbara,
\IEEEauthorrefmark{3}Boston University\\
Email:  jghebre@ncsu.edu, saastha@ucsb.edu, saadu@bu.edu, dahmed@ncsu.edu,\\
gttystah@ncsu.edu, chris@cs.ucsb.edu, vigna@ucsb.edu,
whenck@ncsu.edu, akaprav@ncsu.edu
}
}

\maketitle

\thispagestyle{plain}

\begin{abstract}
Static Application Security Testing (SAST) tools using taint analysis are widely viewed as providing higher-quality vulnerability detection results compared to traditional pattern-based approaches.
However, performing static taint analysis for JavaScript poses two major challenges.
First, JavaScript’s dynamic features complicate data flow extraction required for taint tracking.
Second, npm’s large library ecosystem makes it difficult to identify relevant sources/sinks and establish taint propagation across dependencies.
In this paper, we present \toolname, a multi-agent system that strategically combines the semantic understanding of Large Language Models (LLMs) with traditional static program analysis to extract taint specifications, including sources, sinks, call edges, and library flow summaries tailored to each package. 
Conceptually, \toolname uses static program analysis to calculate a call graph and defers to an LLM to resolve call edges that cannot be resolved statically.
Further, it uses the LLM to classify sources and sinks for a given CWE. 
The resulting taint specification is then provided to a SAST tool, which performs vulnerability analysis.
We integrate \toolname with CodeQL, a state-of-the-art SAST tool, and demonstrate its effectiveness by detecting \testingTP of \testingTotal vulnerabilities previously undetectable by CodeQL.
Furthermore, we find \ossResults novel vulnerabilities in 4 popular npm packages.
In doing so, we demonstrate that LLMs can practically enhance existing static program analysis algorithms, combining the strengths of both symbolic reasoning and semantic understanding for improved vulnerability detection.
\end{abstract}

\section{Introduction}
JavaScript remains the most widely used programming language among developers, with 66\% reporting active use in 2025~\cite{Stack_Overflow_Developer_Survey_2025}. 
The Node Package Manager (npm) registry contains over 3 million JavaScript packages that collectively receive tens of billions of downloads each week~\cite{NPMSTATS}.
Between 2017 and January 2025, npm accumulated 8,237 documented vulnerability reports~\cite{akhavani2025opensourceopenthreats}, the largest of any language ecosystem.
Static Application Security Testing (SAST) tools are central to automated vulnerability discovery~\cite{SAST_POWER}, yet existing tools fail to close this gap for JavaScript.


SAST tools that employ \textit{taint analysis} track the flow of untrusted data through a program to detect when it may reach security-critical operations, making taint analysis essential for discovering non-trivial vulnerabilities~\cite{ANDROMEDA, SAVINGTHEWORLD}.
However, taint analysis depends on data flows derived from call graphs, and JavaScript's dynamic nature fundamentally challenges call graph construction.
A large-scale empirical comparison of static and dynamic techniques found that no existing approach reliably captures JavaScript's call relationships~\cite{static&dynamicCOMP}.
Static tools disagree substantially and most cannot be applied end-to-end on modern Node.js projects; runtime features such as prototype-based inheritance and \texttt{eval()} force static algorithms to conservatively approximate or ignore dynamic behaviors~\cite{MODULARCALLGRAPH,Park_2021}.
Dynamic approaches trace function calls during test execution, but remain bounded by test coverage and require safely executing code that may be malicious or difficult to build reliably~\cite{static&dynamicCOMP}.

Beyond call graph construction, taint analysis requires identifying sources (entry points of untrusted data) and sinks (security-sensitive operations), as well as modeling how taint propagates through third-party library calls. 
A typical npm package transitively depends on roughly 80 libraries~\cite{AverageNPMDeps}, any of which may introduce, transform, or consume tainted data. 
Consequently, popular JavaScript SAST tools such as CodeQL~\cite{CodeQL_Main_Docs}, SonarQube~\cite{SonarQube_Main_Docs}, and Semgrep~\cite{Semgrep_Main_Docs} exclude dependencies from being analyzed by default, relying on manually maintained specifications that cannot keep pace with npm’s scale~\cite{SemGrep_DepIgnore, SonarQube_DepIgnore, CodeQL_Node_Modules}.
While some SAST tools, such as ODGen~\cite{ODGen}, have the ability to analyze the full dependency tree, they suffer greatly from state explosion, making analysis infeasible for large packages.

Brito et al.~\cite{Dataset} curated a dataset of 957 known vulnerabilities in npm packages and compared nine JavaScript SAST tools, including CodeQL and ODGen.
CodeQL out-performed all other SAST tools, 
yet it only detected 31.3\% of the vulnerabilities.
This low recall exposes fundamental gaps in how current SAST tools model JavaScript's dynamic behavior.

We envision combining the strengths of traditional static program analysis with the semantic understanding of LLMs to improve JavaScript vulnerability detection. 
Recent work has begun to couple SAST with LLMs, typically to triage findings or reduce false positives~\cite{SAST-GENIUS, CPGLLM}.
The most relevant prior works, IRIS~\cite{IRIS} and QLPro~\cite{QLPRO}, use LLMs to infer CWE-specific sources and sinks for Java programs and construct CodeQL queries that use these sources and sinks for security analysis. 
However, both share fundamental limitations.
First, both rely on CodeQL's existing call graph to connect identified sources and sinks. 
However, static call graph construction for Java exhibits measurable recall gaps in practice~\cite{JavaCallgraph1, JavaCallgraph2}.
This limitation becomes more pronounced in highly dynamic languages such as JavaScript and Python, where callee resolution is frequently impeded by runtime-dependent features~\cite{Survey_of_Dynamic_Analysis, STRINGDOMAINS, MODULARCALLGRAPH,PythonCallGraph}.
Any unresolved call between source and sink severs the vulnerability path, regardless of how accurately the endpoints are specified.
Second, both IRIS and QLPro invoke the LLM as a stateless classifier for statically extracted APIs.
Due to the inherent limits of LLM context windows, these approaches are restricted to providing signatures, metadata, and at most partial function bodies, preventing the LLM from understanding how each candidate is actually used within the package and whether that usage is potentially vulnerable.
Moreover, neither system can scale to npm's massive API surface: if every API exposed by a package and its transitive dependencies were batched for classification, inference cost and latency would quickly become prohibitive. 
These limitations motivate a fundamentally different approach.

In this paper, we present \toolname, a static-led, LLM-augmented vulnerability analysis tool for JavaScript.
At its core, \toolname treats static analysis as the source of truth, establishing structural facts and flow relations, while surgically invoking LLMs only where static reasoning under-approximates. \toolname addresses limitations of prior approaches by: (1)~repairing broken inter-procedural edges that prevent vulnerability path construction, (2)~exploring code on-demand via targeted LLM invocations rather than processing APIs in isolation, and
(3)~extending analysis across dependency boundaries through direct inspection of library implementations.
Together, these enable holistic analysis along vulnerability paths that prior approaches leave unexplored.


Recent work on LLM-based agents offers an alternative to stateless classification~\cite{RE-ACT}.
Rather than producing a single response from a fixed prompt, agents interleave reasoning with actions by invoking tools (predefined functions), observing results, and iteratively refining their analysis.
\toolname employs three specialized LLM agents, which enhance vulnerability analysis in complementary ways. 
First, the \textbf{\sourceSinkAgentName Agent} identifies potential sources and sinks for vulnerability categories guided by the corresponding MITRE CWE description~\cite{MITRE_CWE}. 
Second, the \textbf{\callgraphAgentName Agent} iteratively resolves call edges that CodeQL cannot determine statically, producing both first-party edges that connect calls to callees within the package and \thirdPartyEdgeName flow summaries for third-party dependencies that conservatively assume taint propagation. 
Third, the \textbf{\flowSummaryAgentName Agent} refines any \thirdPartyEdgeName flow summaries present in vulnerability paths, determining whether the corresponding APIs sanitize or propagate taint, filtering false positives while preserving true vulnerabilities.
The resulting taint specification (sources, sinks, and call edges) is provided to the SAST tool to complete the vulnerability analysis.


We implement a prototype of \toolname for CodeQL, addressing the following three challenges.


\myparagraph{C1: Scaling call graph repair}
Static analysis of JavaScript often leaves many call sites unresolved; real-world npm packages can contain thousands. Exhaustively querying an LLM for every unresolved call is prohibitively expensive in time, compute, and cost. We propose a novel algorithm, \textit{Taint-Informed Callee Resolution (TICR)}, which restricts queries to unresolved calls on potential vulnerability paths for the CWE of interest. 

\myparagraph{C2: Modeling APIs at npm-scale}
CodeQL severs taint flows at library boundaries unless explicitly modeled, because it excludes the code of third-party dependencies from its analysis.
Exhaustively analyzing all transitive third-party dependencies for each package would require prohibitive inference at scale.
\toolname addresses this through \textit{demand-driven dependency modeling}.
\toolname introduces \textit{\thirdPartyEdgeName flow summaries}, conservative placeholders that assume taint propagates through external calls.
The \flowSummaryAgentName Agent then validates only those summaries participating in reported paths, analyzing library implementations to determine whether taint is actually propagated or sanitized.
This deferred validation focuses LLM inference precisely where security results demand it.

\myparagraph{C3: Extracting reusable specifications}
Prior LLM-SAST integrations with CodeQL embed extracted specifications directly into query code, producing single-use artifacts that require recompilation and cannot transfer easily across analyses.
\toolname addresses this through modular specification design using external predicates, enabling iterative refinement without recompilation (essential for TICR).
Moreover, while sources and sinks remain CWE-specific, call graph repairs enhance the foundational inter-procedural dataflow graph shared by all queries, compounding in value across subsequent vulnerability analyses.

\myparagraph{Evaluation \& Results}
%
We evaluated \toolname on 172 vulnerabilities from Brito et al.~\cite{Dataset} with available artifacts, tractable for analysis, yet undetected by CodeQL's taint-tracking queries (Section~\ref{sec:known-vuln-dataset}).
We further reserved \valTotal vulnerabilities for validation.
On the remaining \testingTotal instances, \toolname identified the vulnerability in \testingTP instances, which represents a recall of \textbf{\testingRecall} of vulnerabilities previously undetectable by CodeQL.
Additionally, we used \toolname to study \totalOss open-source npm packages and identified \ossResults previously unknown vulnerabilities.
We have reported these vulnerabilities to the respective package maintainers.

\myparagraph{Contributions}
Static analysis of JavaScript is a long-standing, hard challenge \cite{jsstatic1,jsstatic2,jsstatic3}.
This paper proposes a novel approach of using LLMs to practically enhance existing static program analysis algorithms, combining the strengths of both symbolic reasoning and semantic understanding for improved vulnerability detection.
We demonstrate the effectiveness of this approach with a concrete implementation based on CodeQL and JavaScript, though the concepts can be applied to other SAST tools and programming languages.

\section{Motivation}
\label{sec:motivation}

CodeQL, GitHub's code analysis engine, powers code scanning for over 100,000 repositories and is also available as a VS Code extension or a standalone CLI tool~\cite{CodeQL_Main_Docs,CODEQL_REPOS}. 
CodeQL extracts source code into a relational database and performs taint analysis to detect vulnerabilities through security queries mapped to specific CWEs~\cite{CodeQLCWECoverage}.
Brito et al. curated a dataset of 957 manually verified vulnerabilities from npm advisory reports and evaluated nine JavaScript SAST tools against it~\cite{Dataset}. 
CodeQL outperformed all other tools, yet detected only 31.3\% of the vulnerabilities.
To understand the low performance, we examine an illustrative vulnerability from their dataset that CodeQL misses: \texttt{GHSA-hpr5-wp7c-hh5q}~\cite{GHSA-hpr5-wp7c-hh5q}. This cross-site scripting vulnerability affects a lightweight markdown parser spanning only 320 lines in a single file, \texttt{mrk.js}~\cite{mrk_js}. 
CodeQL fails to find this vulnerability due to multiple layers of indirection that obscure both the sources of untrusted data and the paths to sensitive operations.
The following subsections detail how dynamic behaviors within \texttt{mrk.js} break CodeQL's analysis and prevent vulnerability detection.

\mytitle{Identifying Entry Points}
\label{sec:motivation-sources}
To establish a vulnerability path using taint-tracking, we must first identify the entry point of untrusted data.
As shown in Listing~\ref{code:motivationjs}, the data flow into \texttt{mrk.js} originates at the \texttt{mrk(input)} function parameter on line 5. However, reaching this parameter requires navigating three layers of indirection introduced in its architectural design. The code uses an Immediately Invoked Function Expression \circletext{blue}{1} (IIFE) that executes on module load, exporting a factory function \circletext{customorange}{2} to the module system (Node.js), global object, or browser window. The factory itself returns the \texttt{mrk} function \circletext{customgreen}{4} as a first-class value.

Such multi-layered indirection is common in configurable JavaScript libraries, requiring semantic analysis to track dynamic behavior to identify true taint sources.
CodeQL captures the factory boundary \circletext{red}{3} as a source but does not recognize the returned function's parameter as an entry point, preventing construction of a complete vulnerability path.

\begin{listing}[t]
    \centering
        \includegraphics[
        width=\linewidth,
        trim=0 0 0 0,  
    ]{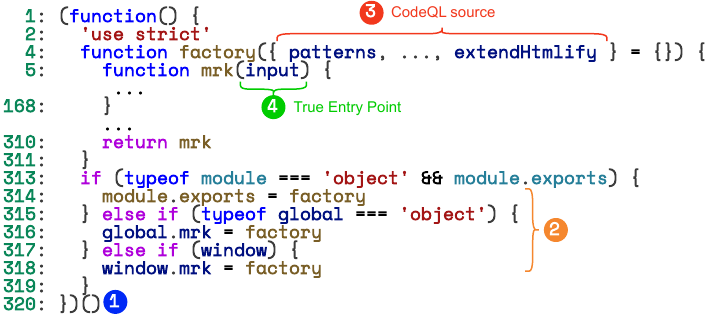}%
\normalsize
\caption{Source of untrusted input in \texttt{mrk.js}}
\label{code:motivationjs}
\end{listing}
\begin{listing}[t]
\centering
    \includegraphics[
    width=\linewidth,
    trim=0 0 0 0,  
]{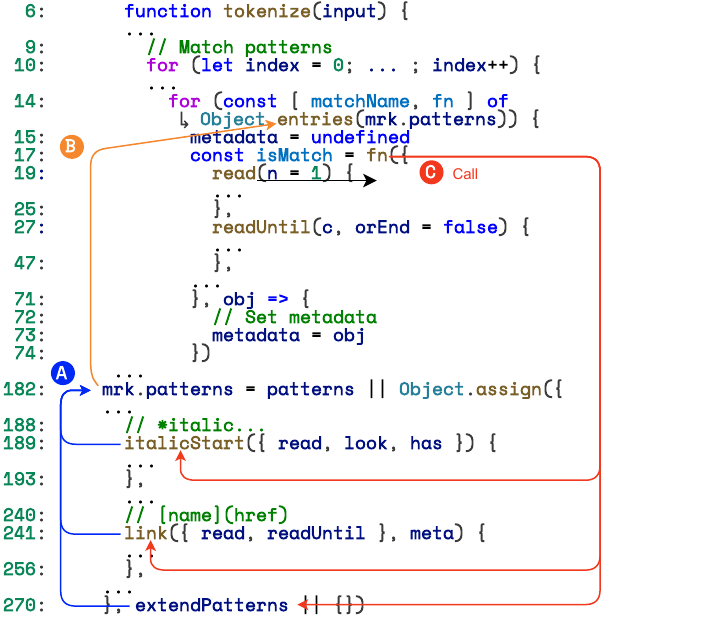}%
\caption{Dynamic dispatch through object composition and reflection in \texttt{mrk.js}}
\label{code:mrk_strat_pattern}
\end{listing}
\mytitle{Resolving Inter-Procedural Data Flow}
Even if CodeQL correctly identified the entry point within \texttt{mrk.js}, tracking how user input flows to the vulnerable operation (sink) requires resolving three calls whose targets are determined by distinct dynamic behaviors.
CodeQL leaves all three unresolved, breaking the vulnerability path.
The following paragraphs step through each of these breaks.

\mysubtitle{Break 1: Enumeration-Based Dispatch} 
The input parameter is passed into a \texttt{tokenize} function that iterates character by character over the markdown string. 
To identify markdown patterns, \texttt{mrk.js} uses a strategy pattern where each markdown element (links, bold, italic) has its own pattern-matching function. 
As shown in Listing~\ref{code:mrk_strat_pattern}, these functions are assembled dynamically: \texttt{Object.assign} \circletext{blue}{A} merges user-provided \texttt{extendPatterns} with default patterns at runtime. 

The tokenizer enumerates these patterns using \texttt{Object.entries(mrk.patterns)} \circletext{customorange}{B}, which iterates over the object's properties at runtime.
For each pattern, the tokenizer invokes \texttt{fn} \circletext{red}{C}, passing a set of parsing utilities. 
This layering of dynamic behaviors prevents CodeQL from statically resolving which pattern functions execute.

\mysubtitle{Break 2: Closure-Based Callback}
When the tokenizer invokes the \texttt{link} pattern function, it passes a callback to the \texttt{meta} parameter \circletext{customlightpurple}{1} that closes over the tokenizer's local \texttt{metadata} variable, as shown in Listing~\ref{code:mrk_callback}.
The \texttt{link} function extracts the URL from markdown and propagates the result by invoking \texttt{meta(\{name, href\})} \circletext{customlightblue}{2} rather than returning it directly.
This callback-based data propagation, common in JavaScript APIs, obscures the data flow path for static analysis.

\begin{listing}[t]
\centering
    \includegraphics[
    width=\linewidth,
    trim=0 0 0 0,  
]{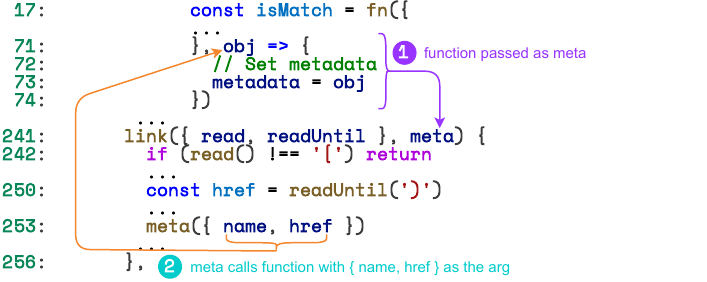}%
\caption{Closure-Based Callback in \texttt{mrk.js}}
\label{code:mrk_callback}
\end{listing}


\mysubtitle{Break 3: Computed Property Dispatch}
After tokenization, the \texttt{html()} method renders tokens by looking up rendering functions using computed property access: \texttt{mrk.htmlify[token.name](token)}.
As shown in Listing~\ref{code:mrk_computed}, the \texttt{htmlify} object \circletext{customlightpurple}{A} is assembled at runtime through \texttt{Object.assign} with user-provided \texttt{extendHtmlify} \circletext{custompink}{C}.
The property key \texttt{token.name} \circletext{customlightblue}{B} is a string determined by which pattern matched during tokenization.
This combination of runtime object composition and computed property dispatch prevents CodeQL from resolving which rendering function receives the tainted data.


\begin{listing}[t]
\centering
    \includegraphics[
    width=\linewidth,
    trim=0 0 0 0,  
]{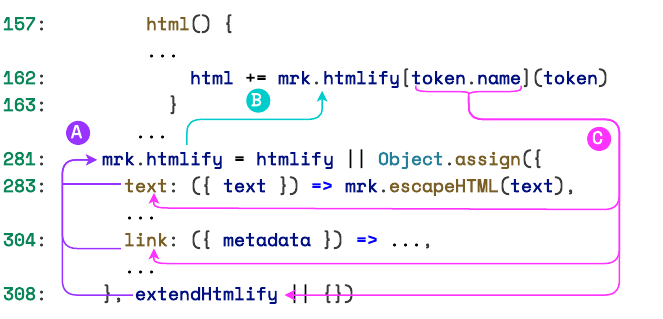}%
\caption{Computed Property Access with Dynamic Dispatch in \texttt{mrk.js}}
\label{code:mrk_computed}
\end{listing}

\begin{listing}[t]
\centering
    \includegraphics[
    width=\linewidth,
    trim=0 0 0 0,  
]{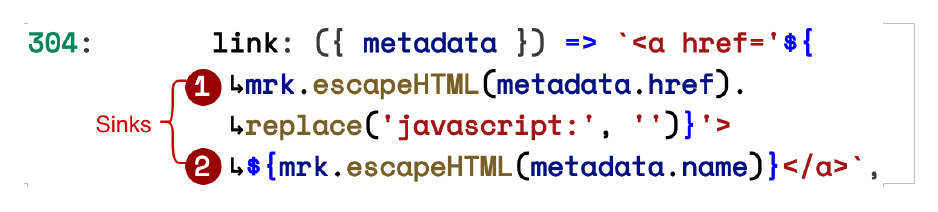}%
\caption{Dangerous Execution in \texttt{mrk.js}}
\label{code:mrk_sink}
\end{listing}

\begin{figure*}[t]
    \centering
    \includegraphics[width=\linewidth]{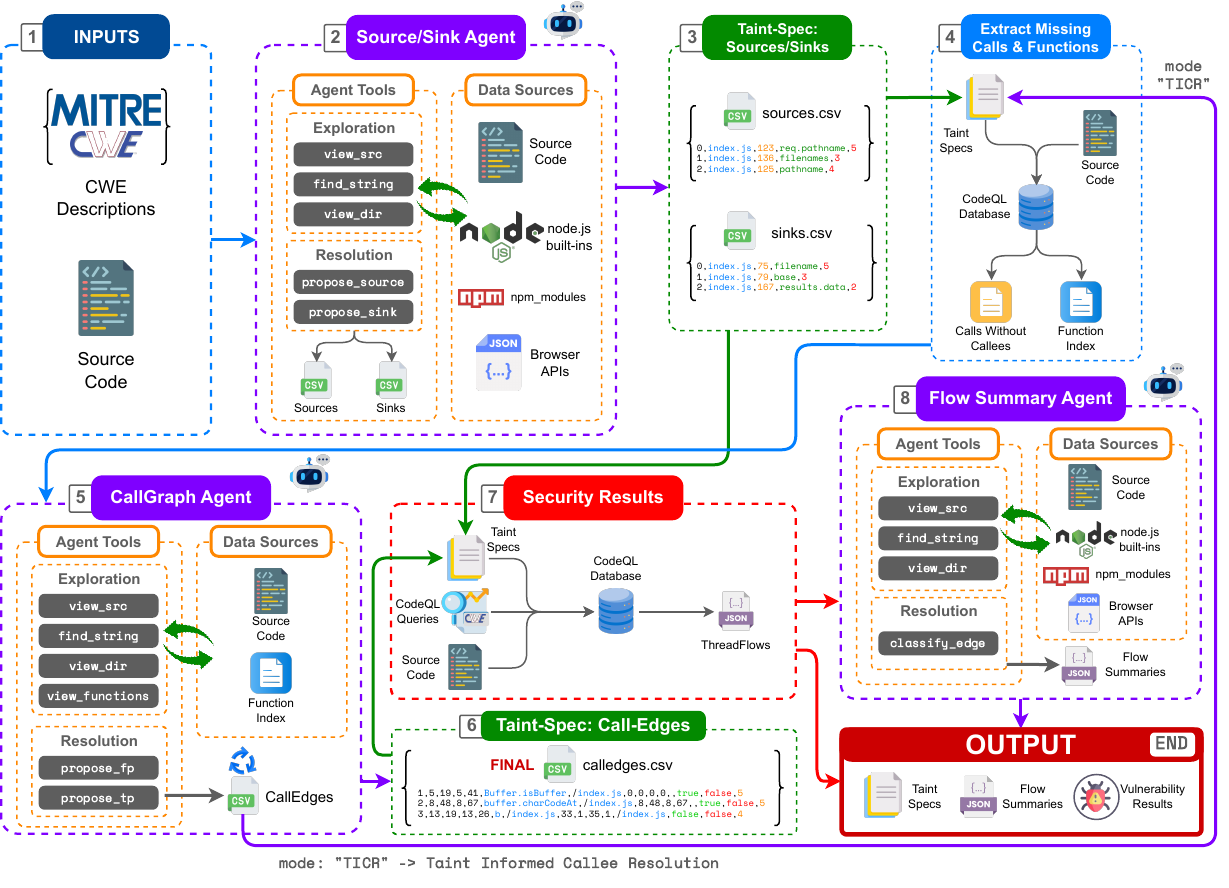}
    \caption{\toolname System Design}
    \label{fig:system_arch}
\end{figure*}



\mytitle{Identifying Dangerous Execution}
\label{sec:motivation-sinks}
Even if taint tracking succeeded through the prior breaks, CodeQL faces a final challenge: identifying the sink.
As shown in Listing~\ref{code:mrk_sink}, the \texttt{link} rendering function constructs HTML using a template literal, interpolating \texttt{metadata.href} \circletext{customdarkred}{1} into an \texttt{href} attribute and \texttt{metadata.name} \circletext{customdarkred}{2} into the link text.

CodeQL cannot recognize this as a sink because the danger is package-specific: no well-known API like \texttt{innerHTML} or \texttt{document.write} is invoked.
Identifying the vulnerability requires understanding that this string will eventually reach an HTML context and that the case-sensitive \texttt{replace('javascript:', '')} sanitization is insufficient.

\mytitle{Broader Context}
These patterns are not unique to \texttt{mrk.js}; they represent common JavaScript development practices found throughout npm packages~\cite{richardsAnalysisDynamicBehavior}. 
The cases we detail point to a broader class of dynamic features that confound static analysis, including prototype manipulation, dynamic imports, and dynamic code generation. 
The fundamental mismatch between JavaScript's runtime flexibility and static analysis assumptions helps explain why even state-of-the-art tools like CodeQL achieve only 31\% recall on known npm package vulnerabilities. Existing LLM-augmented approaches like IRIS and QLPro ~\cite{IRIS, QLPRO} infer sources and sinks but operate on CodeQL's existing call graph; they cannot repair the broken inter-procedural edges that prevent source-to-sink path construction in cases like \texttt{mrk.js}.

Addressing these detection gaps requires a system that can: (1) identify package-specific entry points and dangerous APIs, and (2) semantically reason about dynamic, runtime behavior to resolve call edges. This motivates \toolname's approach: strategically augmenting static analysis with LLM-based reasoning at specific points where dynamic behavior breaks traditional techniques.

\section{Overview}
\label{sec:overview-arch}

\toolname addresses the analysis challenges demonstrated in Section~\ref{sec:motivation} by combining static analysis precision with LLM semantic understanding. To achieve this, \toolname operates over the following program representation extracted from CodeQL:
\begin{description}
    \item[(i)] \textbf{Function Index:} a registry of all function definitions CodeQL successfully extracted.
    \item[(ii)] \textbf{Unresolved Calls:} a set of invocation sites where CodeQL cannot precisely resolve the callee.
    \item[(iii)] \textbf{QL Nodes:} a set of program elements in CodeQL's data flow graph that carry values (expressions, parameters, call sites), connected by edges that model value propagation~\cite{CodeQL_Nodes}.
\end{description}

Agents exchange and refine analysis through a set of artifacts: sources, sinks, call edges, and flow summaries.
The end product is a semantically enriched set of \textit{taint specifications} provided to CodeQL to enhance security analysis. 

\mytitle{Pipeline Flow}
Figure~\ref{fig:system_arch} illustrates \toolname's three-stage modular pipeline. Given a package and CWE \squaretext{custombluedark}{1}, \toolname proceeds as follows:
\begin{enumerate}
    \item The \emph{\sourceSinkAgentName Agent}  \squaretext{custompurple}{2} proposes sources and sinks specific to both the package and the CWE, binds them to  QL Nodes, and exports them as taint-specification facts \squaretext{customgreen}{3}.
    \item The \emph{\callgraphAgentName Agent} \squaretext{custompurple}{5} consumes the unresolved call set  \squaretext{custombluelight}{4} and emits two kinds of edges, bound to QL Nodes~\squaretext{customgreen}{6}: \textbf{(a)} First-party call edges that connect call sites to callees that reside in the package's source code. \textbf{(b)} \ThirdPartyEdgeName flow summaries for third-party calls: conservative placeholders that initially assume taint propagation (and are refined later). 
    \item The \emph{\flowSummaryAgentName Agent} \squaretext{custompurple}{8} runs after security queries have executed with the learned taint-specs on CodeQL \squaretext{red}{7}. It validates only those \thirdPartyEdgeName summaries appearing in reported paths, determining whether each third-party call propagates or sanitizes taint. This refinement produces precise flow summaries, enabling accurate alerts despite CodeQL's exclusion of dependency code.

    \item The \textit{Final Output} \squaretext{customred}{END} of \toolname is a taint specification comprising validated sources, sinks, and call edges, which CodeQL uses to produce vulnerability detection results.
\end{enumerate}

\toolname optionally accepts existing taint-specifications as input, enabling analysis of additional CWEs on the same package to leverage previous repairs.
Call graph improvements thus compound across subsequent analyses (\textbf{C3}).

\myparagraph{Shared tooling}
All agents inherit a minimal, read-only \textit{traversal toolkit} (\texttt{view\_src}, \texttt{find\_string}, \texttt{view\_dir}) to investigate the codebase consistently; each agent adds task-specific tools aligned with its responsibility (see Appendix~\ref{sec:agent-implementation}).

\myparagraph{Multi-run aggregation}
Due to the stochastic nature of LLM reasoning, identical prompts can yield different answers across runs~\cite{LLMSTABILITY}. We incorporate agent-specific aggregation strategies, executing each analysis multiple times and pooling results to produce stable outcomes~\cite{aggregation_LLMs, Ensemble_2025, VULN_REPAIR_2025} (see Appendix~\ref{sec:agent-implementation}).

\myparagraph{Validation}
\toolname verifies that every proposed fact binds to CodeQL Data Flow nodes. 
Items that fail to bind are rejected; this gating ensures agents only produce specifications that can be represented and used within CodeQL.

\myparagraph{Confidence calibration}  
LLMs exhibit systematic overconfidence, clustering self-reported certainty in narrow high bands regardless of actual accuracy~\cite{LLM_CONFIDENCE}. 
We address this through calibrated rubrics that anchor each score level to explicit, observable criteria (Appendix~\ref{appendix:confidence}).

\myparagraph{CWE Context}  
\toolname includes pre-parsed MITRE CWE entries~\cite{MITRE_CWE} for each of CodeQL's 37 CWEs supporting taint-analysis queries.
Each entry contains a structured representation (ID, name, description, and common consequences) that informs agents what constitutes untrusted data and vulnerable operations for each vulnerability class.

This section has treated agents as black boxes that consume and produce taint-specification artifacts. Section~\ref{sec:semtaint} formalizes how we craft and validate these taint-specifications and details how each agent combines LLM-based semantic reasoning to extract precise analysis facts from JavaScript codebases.

\section{\toolname}
\label{sec:semtaint}

In this section, we formalize \toolname's design. We begin by defining the structure of taint specifications and their bindings to QL Nodes. Then we illustrate how each agent extracts and validates these specifications.

\subsection{Taint Specifications}%
\label{sec:taint-specs}

Taint specifications are used by taint-tracking static analysis tools to enumerate entry points that introduce untrusted data as \textit{sources}, APIs that perform security-sensitive operations as \textit{sinks}, \textit{propagation rules} that describe how taint flows through data dependencies, and \textit{sanitizers} that neutralize tainted data~\cite{TASER,almashfiStaticTaintAnalysis2021}. 
By overlaying security semantics onto the analyzer's generic data flow graph, these specifications map reachability into exploitability.
Taint specifications are often organized around the Common Weakness Enumeration (CWE)~\cite{MITRE_CWE, piskachev}, which defines over 900 weakness types; each CWE represents a distinct security weakness with characteristic sources, sinks, and propagation semantics. 
\toolname enables CodeQL to discover previously undetectable vulnerabilities by generating CWE-specific sources and sinks, as well as selectively repairing call graph edges along potential vulnerability paths.

\myparagraph{Locations}
\toolname grounds agent outputs with \emph{location tuples} that identify specific syntactic locations in a codebase.
We use two location-tuple formats that we will formalize here.

Let $\mathcal{F}$ be the set of file paths within a package, $\mathbb{N}$ the natural numbers,
and $\Sigma^{*}$ the set of finite strings.

An \emph{endpoint location} records a snippet occurrence using its start coordinate and a snippet string:
\begin{equation}\label{eq:endpoint-location}
\begin{aligned}
\mathcal{L}_{\mathrm{E}} &\coloneqq \mathcal{F}\times\mathbb{N}\times\mathbb{N}\times\Sigma^{*},\\
\tau &= \langle f,\ell,\mathrm{col},v\rangle \in \mathcal{L}_{\mathrm{E}}.
\end{aligned}
\end{equation}
where $f\in\mathcal{F}$ is a file path, $(\ell,\mathrm{col})$ is the start line/column,
and $v\in\Sigma^{*}$ is a short code snippet string.

A \emph{span location} identifies a contiguous region of code by its start and end coordinates:
\begin{equation}\label{eq:span-location}
\begin{aligned}
\mathcal{L}_{\mathrm{S}} &\coloneqq \mathcal{F}\times\mathbb{N}^{4},\\
\lambda &= \langle f,\ell_s,\mathrm{col}_s,\ell_e,\mathrm{col}_e\rangle \in \mathcal{L}_{\mathrm{S}}.
\end{aligned}
\end{equation}
where $f\in\mathcal{F}$, $(\ell_s,\mathrm{col}_s)$ is the start line/column, and
$(\ell_e,\mathrm{col}_e)$ is the end line/column.

The \sourceSinkAgentName Agent uses endpoint locations for sources/sinks, while the \callgraphAgentName Agent uses span locations for invocations and callee functions.
We use span locations for call edges because function definitions span multiple lines and may nest; the agent selects targets from a pre-built function index, and coordinates provide reliable binding.
Sources and sinks are typically concise expressions that the LLM identifies directly as snippet strings.

\subsubsection{Sources}
\label{sec:taint-specs-sources}
A \textit{source} is an entry point of untrusted data into an application.
In practice, sources can be obscured by package-specific architectural patterns.
The true entry point of untrusted data may be hidden behind layers of indirection (Section~\ref{sec:motivation-sources}) or exposed through third-party APIs.

The \sourceSinkAgentName Agent (Figure~\ref{fig:system_arch}~\squaretext{custompurple}{2}) proposes sources as endpoint locations
$\tau\in\mathcal{L}_{\mathrm{E}}$ (Eq.~\ref{eq:endpoint-location}), excluding locations within test directories as these are not relevant to vulnerability detection.

Let $\mathcal{N}_{\mathrm{QL}}$ denote the set of QL nodes.
We resolve endpoint locations to QL nodes via the partial function
\[
  \mathsf{qlnode}:\mathcal{L}_{\mathrm{E}}\rightharpoonup \mathcal{N}_{\mathrm{QL}}.
\]
$\mathsf{qlnode}(\tau)$ is defined iff there exists
$n\in\mathcal{N}_{\mathrm{QL}}$ such that
\begin{equation}\label{eq:qlnode}
\begin{aligned}
\text{file}(n) = f, \quad &\text{startLine}(n) = \ell, \\
\text{startCol}(n) = \mathrm{col}, \quad &\text{toString}(n) = v \,.
\end{aligned}
\end{equation}
In that case, we set $\mathsf{qlnode}(\tau)=n$, otherwise it is undefined. We write $\mathsf{qlnode}(\tau)\downarrow$ to denote that $\mathsf{qlnode}$ is defined at $\tau$.

\myparagraph{Source facts}
Let $C=\{1,2,3,4,5\}$.
The \sourceSinkAgentName Agent extracts source facts of the form $\langle id,\tau,c\rangle$, collected in
\[
  S \subseteq \mathbb{Z}\times \mathcal{L}_{\mathrm{E}} \times C,
\]
where $id$ is a unique identifier and $c\in C$ is a confidence score.
The set of QL nodes corresponding to successfully bound sources is
\begin{equation}\label{eq:bound_source}
S^{\uparrow}\ \coloneqq\
\left\{\,\mathsf{qlnode}(\tau)\ \middle|\ \langle id,\tau,c\rangle\in S\ \wedge\ \mathsf{qlnode}(\tau)\downarrow\,\right\}.
\end{equation}

\subsubsection{Sinks}
\label{sec:taint-specs-sinks}
A \textit{sink} is a security-sensitive operation that may be exploited when reached by tainted data.
Classifying sinks requires understanding both package-specific patterns that route data to dangerous operations (Section~\ref{sec:motivation-sinks}) and third-party APIs that perform security-relevant execution.

The \sourceSinkAgentName Agent (Figure~\ref{fig:system_arch}~\squaretext{custompurple}{2}) proposes sinks as \emph{endpoint locations} $\tau\in\mathcal{L}_{\mathrm{E}}$ (Eq.~\ref{eq:endpoint-location}), subject to the same test-directory exclusion as sources.
We resolve endpoint locations to QL nodes using the partial function $\mathsf{qlnode}$ (Eq.~\ref{eq:qlnode}).

\myparagraph{Sink facts}
Let $C=\{1,2,3,4,5\}$.
The agent emits sink facts of the form $\langle id,\tau,c\rangle$, collected in
\[
  K \subseteq \mathbb{Z}\times \mathcal{L}_{\mathrm{E}} \times C,
\]
where $id$ is a unique identifier and $c\in C$ is a confidence score.
The set of QL nodes corresponding to successfully bound sinks is
\begin{equation}\label{eq:bound_sink}
K^{\uparrow}\ \coloneqq\
\left\{\,\mathsf{qlnode}(\tau)\ \middle|\ \langle id,\tau,c\rangle\in K \ \wedge\ \mathsf{qlnode}(\tau)\downarrow\,\right\}.
\end{equation}

\subsubsection{Propagation Rules}
\label{sec:InterproceduralFlows}

A \textit{propagation rule} specifies how taint on an operation's inputs transfers to its outputs~\cite{PropRules1}.
Taint analysis tools commonly distinguish \emph{intra-procedural} rules, which transfer taint through expressions and statements within a function, from \emph{inter-procedural} rules, which transfer taint across function calls and returns~\cite{CodeQLPropRules, PropRules1, INSPECTJS}.
\toolname{} leaves intra-procedural propagation rules intact, focusing instead on recovering missing inter-procedural call edges.

\myparagraph{Inter-procedural data flow}
\toolname{} enhances inter-procedural data flow by extracting \textit{call edges}: mappings from invocations (call sites) to their target callee functions.
We distinguish two classes:
(1)~\textit{first-party edges}, which connect call sites to callees defined within the analyzed package; and
(2)~\textit{third-party edges}, which connect call sites to callees outside the package.

\toolname distinguishes between structural and implementation gaps.
For first-party calls, CodeQL can derive propagation semantics from function bodies but cannot determine which function a dynamic call site invokes.
Recovering these missing edges enables CodeQL's existing analysis to traverse complete paths automatically.
These repairs enhance the foundational data flow graph shared by all queries, benefiting every vulnerability type rather than targeting a single CWE.

Third-party calls lack implementations entirely due to CodeQL excluding \texttt{node\_modules} analysis by default~\cite{CodeQL_Node_Modules}.
Without callee code in the database, propagation behavior must be specified abstractly via \textit{flow summaries}: specifications that describe whether taint propagates from function arguments to return values without modeling internal data flow paths~\cite{TASER, INSPECTJS}.
Unlike universal call graph repairs, these summaries are CWE-specific: HTML encoding sanitizes XSS (Cross-Site Scripting) but not command injection.

The \callgraphAgentName Agent (Figure~\ref{fig:system_arch}~\squaretext{custompurple}{5}) links unresolved call sites to target functions from a function index, producing call edges as pairs of \emph{span locations} $\lambda\in\mathcal{L}_{\mathrm{S}}$ (Equation~\ref{eq:span-location}).

First-party edges require both spans to bind to QL nodes; third-party edges require only the call site to bind.

Let $\mathcal{I}_{\mathrm{QL}}$ denote the set of \texttt{Invocation} QL nodes and $\mathcal{F}_{\mathrm{QL}}$ the set of \texttt{Function} QL nodes.
We resolve span locations to these QL nodes via the partial function
\[
  \mathsf{qlinvoke}: \mathcal{L}_{\mathrm{S}} \rightharpoonup \mathcal{I}_{\mathrm{QL}}.
\]
$\mathsf{qlinvoke}(\lambda)$ is defined iff there exists $n \in \mathcal{I}_{\mathrm{QL}}$ such that
\begin{equation}\label{eq:qlinvoke}
\begin{aligned}
\text{file}(n) = f, \quad &\text{startLine}(n) = \ell_s, \quad \text{startCol}(n) = col_s, \\
&\text{endLine}(n) = \ell_e, \quad \text{endCol}(n) = col_e \, .
\end{aligned}
\end{equation}
In that case, we set $\mathsf{qlinvoke}(\lambda) = n$.
We define an analogous partial function for callee functions:
\begin{equation}\label{eq:qlcallee}
  \mathsf{qlcallee}: \mathcal{L}_{\mathrm{S}} \rightharpoonup \mathcal{F}_{\mathrm{QL}}.
\end{equation}

\myparagraph{Call edge facts}
Let $B = \{\mathsf{first}, \mathsf{third}\}$ and $C = \{1,2,3,4,5\}$.
The \callgraphAgentName Agent extracts call edge facts of the form $\langle id, \lambda_{\mathrm{call}}, \lambda_{\mathrm{target}}, \beta, c \rangle$, collected in
\[
  E \subseteq \mathbb{Z} \times \mathcal{L}_{\mathrm{S}} \times \mathcal{L}_{\mathrm{S}} \times B \times C,
\]
where $id$ is a unique identifier, $\lambda_{\mathrm{call}}$ and $\lambda_{\mathrm{target}}$ are span locations for the call site and target function, $\beta \in B$ records whether the callee is first- or third-party, and $c \in C$ is a confidence score.

\myparagraph{First-party edges}
For first-party edges, $\beta = \mathsf{first}$ and both endpoints must bind to QL nodes.
The set of successfully bound first-party edges is
\begin{equation}\label{eq:bound_fp_edge}
\begin{aligned}
E^{\uparrow}_{\mathsf{first}} \coloneqq \Big\{\, 
  &\bigl(\mathsf{qlinvoke}(\lambda_{\mathrm{c}}),\, \mathsf{qlcallee}(\lambda_{\mathrm{t}})\bigr) \;\Big| \\[2pt]
  &\langle id, \lambda_{\mathrm{c}}, \lambda_{\mathrm{t}}, \mathsf{first}, c \rangle \in E \ \wedge \\[2pt]
  &\mathsf{qlinvoke}(\lambda_{\mathrm{c}})\!\downarrow \ \wedge\ \mathsf{qlcallee}(\lambda_{\mathrm{t}})\!\downarrow
\,\Big\}
\end{aligned}
\end{equation}
where $\lambda_{\mathrm{c}}$ and $\lambda_{\mathrm{t}}$ abbreviate the call-site and target spans.
Each pair $(i, t) \in E^{\uparrow}_{\mathsf{first}}$ is asserted as an additional inter-procedural edge in CodeQL's data flow graph.

\myparagraph{Third-party edges}
For third-party edges, $\beta = \mathsf{third}$ and the callee implementation lies outside the database.
We retain only the call site:
\begin{equation}\label{eq:bound_tp_edge}
\begin{aligned}
E^{\uparrow}_{\mathsf{third}} \coloneqq \Big\{\,
  &\mathsf{qlinvoke}(\lambda_{\mathrm{c}}) \;\Big| \\[2pt]
  &\langle id, \lambda_{\mathrm{c}}, \lambda_{\mathrm{t}}, \mathsf{third}, c \rangle \in E \ \wedge\ 
  \mathsf{qlinvoke}(\lambda_{\mathrm{c}})\!\downarrow
\,\Big\}
\end{aligned}
\end{equation}



\myparagraph{\ThirdPartyEdgeName flow summaries}
Third-party unresolved calls pose a distinct challenge: CodeQL excludes dependency code by default, thus no target callee representation exists in the database.
Rather than importing entire transitive dependency trees during call graph repair, \toolname treats each $i \in E^{\uparrow}_{\mathsf{third}}$ as a \textit{\thirdPartyEdgeName flow summary}: a conservative placeholder that assumes taint propagates through the external call pending validation.

This separation deliberately decouples structural call graph repair (CWE-agnostic and reusable) from modeling the taint effects of dependencies (CWE-specific) through \textit{demand-driven dependency modeling} (\textbf{C2}), which yields two efficiencies.
\textbf{(1)} \emph{Efficient third-party identification:} CodeQL's API graph~\cite{CodeQLAPIGraph} can often determine that an unresolved call crosses the package boundary by tracing access paths from module imports.
\toolname uses custom queries leveraging this API graph to auto-classify nearly half of unresolved calls as third-party without LLM inference.
When the API graph is incomplete, the \callgraphAgentName Agent can still identify third-party calls, emitting candidate summaries without analyzing library internals.
\textbf{(2)} \emph{Demand-driven validation:} Analyzing taint propagation through third-party calls within the \callgraphAgentName Agent would require loading external repositories, bloating the agent's context even though most third-party calls never appear on a source--sink path.
Instead, candidate summaries are validated only when they participate in reported vulnerability paths through the specialized \flowSummaryAgentName Agent (Figure~\ref{fig:system_arch}~\squaretext{custompurple}{8}).
This task isolation keeps both agents focused on their specialized roles, consistent with findings that modular agent designs outperform monolithic alternatives~\cite{CVEGENIE, AGENT_ROLES1}: the \callgraphAgentName Agent resolves call targets without navigating library codebases, while the \flowSummaryAgentName Agent models propagation semantics without resolving call sites.
Section~\ref{sec:RQ3} quantifies the payoff of this demand-driven design, showing that on-demand validation avoids the vast majority of third-party library analyses while preserving detection capability.

\subsubsection{Sanitizers}
\label{sec:sanitizers}

A \emph{sanitizer} is a function or conditional check whose output is considered safe regardless of whether its input was tainted~\cite{Sanitizers}.
SAST tools, like CodeQL, terminate taint flows that pass through recognized sanitizers, avoiding false positives when untrusted input is handled safely~\cite{CodeQLPropRules}.
\toolname relies on CodeQL's existing sanitizer infrastructure for standard taint analysis.
Our contribution to handling sanitization occurs through the \flowSummaryAgentName Agent (Figure~\ref{fig:system_arch}~\squaretext{custompurple}{8}), which validates candidate flow summaries appearing in vulnerability paths, determining whether each library call sanitizes or propagates taint for the target CWE.

\subsection{\sourceSinkAgentName Agent}
\label{sec:source_sink_agent}
CodeQL's taint-tracking queries require accurate source and sink specifications to construct vulnerability paths.
Although CodeQL provides default specifications for standard APIs, these cannot capture package-specific architectures such as those in Section~\ref{sec:motivation}, requiring semantic extraction tailored to each package under analysis.

For this extraction, we employ the \sourceSinkAgentName Agent, which uses CWE-specific context (Section~\ref{sec:overview-arch}) to identify package-specific sources and sinks.
In addition to package source code, the \sourceSinkAgentName Agent has access to three code repositories for analysis:
\begin{enumerate}[itemsep=0pt,topsep=0pt,leftmargin=*]
    \item \textbf{node\_modules}: The installed npm dependencies for each package under analysis, which enable the agent to read actual library source code rather than relying on documentation or heuristics. 
    \item \textbf{Node.js built-in modules}: Core modules (e.g., \texttt{child\_process}, \texttt{fs}, \texttt{http}) that ship with Node.js~\cite{NodeBuiltinDocs} and define APIs available to npm packages without explicit installation.
    \item \textbf{Browser API index}: An indexed JSON database of 1,175 browser API objects, which enables reasoning about browser-specific functions that may not have accessible source code. Extracted from the Chromium codebase and provided in VisibleV8 releases \cite{VV8, VisibleV8Release}.
\end{enumerate}

Given a package's source code and CWE number, the \sourceSinkAgentName Agent receives the corresponding structured CWE context and selectively explores the codebase and third-party dependencies. The agent searches for program elements matching the CWE's characteristic patterns: potential entry points of untrusted data for sources and security-sensitive operations for sinks. The LLM applies semantic understanding to recognize both standard API patterns and package-specific variants. For instance, given CWE-79 (Cross-site scripting), the agent not only identifies standard sources like \texttt{req.query.input}, but also the package-specific \texttt{mrk(input)} parameter hidden behind the factory closure in {\tt mrk.js}. Appendix~\ref{sec:source-sink-agent-walkthrough} traces this process in detail.

For each candidate source or sink, the agent proposes Source $S$ and Sink $K$ facts (Sections~\ref{sec:taint-specs-sources}~\&~\ref{sec:taint-specs-sinks}) with the \texttt{propose\_source} and \texttt{propose\_sink} tools. These sources and sinks are then provided to the \callgraphAgentName Agent to assist in its discovery of security-relevant unresolved calls.

\subsection{CallGraph Agent}
\label{sec:CallGraph}
Once CWE-specific sources and sinks are identified, CodeQL's vulnerability analysis hinges on its ability to connect them through its data flow graph.
In JavaScript, many dynamic calls remain unresolved in CodeQL's call graph, severing taint propagation at function boundaries and third-party API calls.
The \callgraphAgentName Agent repairs these broken edges, establishing the inter-procedural connections needed to construct vulnerability paths.

\subsubsection{Unresolved calls}
\label{sec:callgraph-unresolved-calls}
The \callgraphAgentName Agent targets calls left unresolved within CodeQL's existing call graph.
In CodeQL's JavaScript data flow library, each call site is an Invocation QL Node, and the \texttt{getACallee} predicate enumerates its potential callee functions at different imprecision levels~\cite{CodeQL_Get_A_CALLEE_2025}. 
At precision 0, callees are derived without relying on global flow heuristics, and thus are considered precise.
Therefore, we define an unresolved call as an invocation with no resolved callee at precision 0, excluding calls within test directories as these are not relevant to vulnerability detection.
Formally, letting $\mathcal{I}_{\text{QL}}$ be the set of Invocation QL Nodes and $\mathcal{F}_{\text{QL}}$ the set of Function QL Nodes, the set of unresolved calls is
\begin{equation}\label{eq:missing_calls_raw}
\begin{aligned}
\mathcal{M}_{\text{raw}}
  = \bigl\{\, i \in \mathcal{I}_{\text{QL}} \;\bigm|\;
  & \nexists\, t \in \mathcal{F}_{\text{QL}} \\
  & \text{s.t. } i.\texttt{getACallee}(0)=t \,\bigr\}.
\end{aligned}
\end{equation}
We filter $\mathcal{M}_{\text{raw}}$ to remove calls for which CodeQL provides internal flow summary rules that model taint propagation without requiring callee resolution.
The resulting set of unresolved calls is
\begin{equation}\label{eq:missing_calls}
\mathcal{M} = \mathcal{M}_{\text{raw}} \setminus \mathcal{F_{\text{s}}}
\end{equation}
where $\mathcal{F_{\text{s}}}$ denotes calls covered by internal flow summaries.
All subsequent analysis operates on $\mathcal{M}$.

\mytitle{Finding Unresolved Calls}
For each unresolved call, we extract identifying information (Figure~\ref{fig:system_arch}~\squaretext{custombluelight}{4}): (1) the invocation's file path; (2) line and column positions; (3) the callee name (when available from the AST); and (4) whether the call appears to target first-party or third-party code based on CodeQL's API Graph.

\myparagraph{Scale of the problem} 
To quantify the prevalence of unresolved calls in JavaScript, we analyzed 938 of the 957 packages in the Brito et al.\ dataset~\cite{Dataset}; the remainder lacked available artifacts (Section~\ref{sec:known-vuln-dataset}).
Of 2,849,543 total call sites, CodeQL resolves only 1,147,393 (40.3\%), leaving 1,702,150 (59.7\%) unresolved ($\mathcal{M}_{\text{raw}}$).
Filtering out calls covered by internal flow summaries ($\mathcal{F_{\text{s}}}$) reduces this to 1,423,995 (49.9\%) unresolved calls ($\mathcal{M}$).
In other words, CodeQL cannot confidently approximate callee resolution for nearly half of function calls in JavaScript applications, even after accounting for internal flow summaries.
Due to the dynamic nature of JavaScript, CodeQL deliberately uses a simple call graph algorithm when constructing its data flow graph~\cite{CodeQL_Node_Modules}. 
More aggressive algorithms can increase recall, but on realistic Node.js projects they often fail to complete, miss multi-file calls, or introduce spurious call targets~\cite{static&dynamicCOMP}.


Cost and time constraints incurred in resolving all callees motivate \toolname's optimized call edge discovery strategy. Rather than attempting to resolve every missing callee in the codebase, \toolname identifies the specific set of unresolved calls that enable taint flow from sources to sinks for a given CWE, and directs the LLM's reasoning exclusively to those high-value targets. 

\subsubsection{Taint-Informed Callee Resolution (TICR)}
\label{sec:ticr}
An \textit{exhaustive} approach would attempt to resolve all unresolved calls.
However, for vulnerability detection, only a fraction of these calls matter: those lying on potential taint paths between sources and sinks.
For instance, an unresolved call along the data flow from user input to a database query may block detection of an injection vulnerability, while a call outside any source-to-sink path has no effect on the analysis.
Rather than resolving all unresolved calls, we leverage CodeQL's taint-tracking infrastructure to identify only those on potential vulnerability paths, significantly reducing the agent's workload.

To address \textbf{C1}, \toolname employs \textit{Taint-Informed Callee Resolution (TICR)}: a bidirectional algorithm that traces taint flow from sources toward unresolved calls and from unresolved calls toward sinks.
An unresolved call becomes a resolution candidate if and only if it is taint-reachable in at least one direction (i.e., it could plausibly lie on a source-to-sink path).
As Figure~\ref{fig:taint-vuln-path} illustrates, TICR targets only unresolved calls on potential vulnerability paths (red), ignoring the majority that cannot affect taint analysis.
The \callgraphAgentName Agent thus avoids expensive inference on calls irrelevant to security analysis.

Formally, let $G = (\mathcal{N}_{\text{QL}}, \mathcal{E}_{\text{flow}})$ denote CodeQL's taint flow graph, where $\mathcal{N}_{\mathrm{QL}}$ represents the set of QL nodes and $\mathcal{E}_{\text{flow}}$ represents taint-preserving edges.
We define the reachability relation $\leadsto$ where $n_1 \leadsto n_2$ if there exists a path in $G$ from $n_1$ to $n_2$ along edges in $\mathcal{E}_{\text{flow}}$.
Using the bound source set $S^{\!\uparrow}$ and sink set $K^{\!\uparrow}$ from Equations~\eqref{eq:bound_source} and~\eqref{eq:bound_sink}, TICR identifies two subsets of the unresolved call set $\mathcal{M}$ (Equation~\eqref{eq:missing_calls}):
\begin{equation}\label{eq:ticr_src_to_brk}
\mathcal{M}_{\text{src→brk}} = \bigl\{\, i \in \mathcal{M} \;\bigm|\; \exists\, s \in S^{\!\uparrow} \text{ s.t. } s \leadsto i \,\bigr\}
\end{equation}
\begin{equation}\label{eq:ticr_brk_to_snk}
\mathcal{M}_{\text{brk→snk}} = \bigl\{\, i \in \mathcal{M} \;\bigm|\; \exists\, k \in K^{\!\uparrow} \text{ s.t. } i \leadsto k \,\bigr\}
\end{equation}

Intuitively, $\mathcal{M}_{\text{src→brk}}$ captures calls where taint flows \emph{into} the unresolved invocation, while $\mathcal{M}_{\text{brk→snk}}$ captures calls where taint can flow \emph{out of} the invocation toward a sink.
The set of \emph{security-relevant} unresolved calls is their union:
\begin{equation}\label{eq:ticr_union}
\mathcal{M}_{\text{TICR}} = \mathcal{M}_{\text{src→brk}} \cup \mathcal{M}_{\text{brk→snk}}
\end{equation}
Appendix~\ref{appendix:ticr-implementation} illustrates the necessity of this bidirectional approach.

\myparagraph{Extended Taint Semantics for JavaScript}
JavaScript's dynamic features create not only inter-procedural data flow gaps but also intra-procedural gaps~\cite{VALUEPARITION, CORRTRACKING}.
Moreover, CodeQL's taint semantics are not designed for call graph resolution: taint does not propagate into or out of a call whose callee is unresolved, severing potential vulnerability paths.
To address these gaps, we extend CodeQL's taint-flow graph $G$ with four conservative flow rules tailored to JavaScript's dynamic patterns, enabling TICR to trace taint to and from unresolved calls that would otherwise be unreachable.

Figure~\ref{fig:extended_flow_rules} presents these rules, where $n \rightarrow m$ denotes a flow edge added to $\mathcal{E}_{\text{flow}}$ by \toolname.
\begin{itemize}
    \item \textsc{Param} propagates taint from arguments into unresolved calls.
    \item \textsc{Object} propagates taint from property writes to containing objects.
    \item \textsc{Func-Obj} handles first-class functions by connecting function parameters, returns, and the function itself to their containing object.
    \item \textsc{Method} propagates taint from an object to unresolved method calls on its descendants.
\end{itemize}
\textit{Notation}: $\mathsf{args}(i)$ denotes invocation arguments; $\mathsf{props}(o)$ object properties; $\mathsf{rhs}(w)$ and $\mathsf{base}(w)$ the right-hand side and base of a write; $\mathsf{params}(f)$ and $\mathsf{rets}(f)$ function parameters and returns; $\mathsf{desc}^*(o)$ transitive property access from $o$.

\begin{figure}[t]
\small
\setlength{\abovedisplayskip}{4pt}
\setlength{\belowdisplayskip}{4pt}
\begin{align*}
\textsc{Param}:\quad
& \frac{n \in \mathsf{args}(i) \quad i \in \mathcal{M}}{n \rightarrow i} 
\\[1em]
\textsc{Object}:\quad
& \frac{n \in \mathsf{props}(o)}{n \rightarrow o}
\quad
\frac{n = \mathsf{rhs}(w) \quad o = \mathsf{base}(w)}{n \rightarrow o}
\\[1em]
\textsc{Func-Obj}:\quad
& \frac{f \in \mathsf{props}(o) \quad n \in \mathsf{params}(f) \cup \mathsf{rets}(f) \cup \{f\}}{n \rightarrow o}
\\[1em]
\textsc{Method}:\quad
& \frac{i \in \mathsf{desc}^*(o) \quad i \in \mathcal{M}}{o \rightarrow i}
\end{align*}
\vspace{-6pt}
\caption{Extended taint propagation rules.}
\label{fig:extended_flow_rules}
\end{figure}

\subsubsection{Agent-Based Call Resolution}

The \callgraphAgentName Agent (Figure~\ref{fig:system_arch}~\squaretext{custompurple}{5}) performs focused, per-call analysis on the security-relevant unresolved call set.
For each call, the agent receives identifying context: file path, line and column location, and callee name (when available from the AST).
The agent systematically traces from the call site to candidate function definitions, documenting each step of the resolution path.

For first-party resolutions where the target lies within the analyzed package, the agent produces a granular trace of its reasoning steps.
This trace serves two purposes: (1) it enforces documentation of intermediate reasoning (chain-of-thought), improving LLM performance on multi-step tasks~\cite{LLM_COT}, and (2) it provides an auditable record for manual examination.
Appendix~\ref{appendix:callgraph-agent} demonstrates this process for the dynamic dispatch patterns in {\tt mrk.js} from Section~\ref{sec:motivation}.
For third-party resolutions, the agent instead returns structured metadata identifying the target library (package name, version, entry point).

By default, the agent processes $\mathcal{M}_{\text{TICR}}$ (Equation~\ref{eq:ticr_union}) via iterative refinement.
In iteration $t$, the agent resolves a subset of calls, producing new call graph edges.
These edges enable CodeQL to propagate taint further, potentially exposing additional unresolved calls along source-to-sink paths that seed the next iteration.
The process terminates when no new security-relevant calls are discovered or a configurable iteration limit is reached.

TICR's failure-handling mechanism is critical to precision: when the agent cannot confidently resolve a call due to insufficient evidence, ambiguity, or missing function index entries, it marks the call as \emph{ignored} rather than proposing a speculative edge.
Ignored calls are excluded from subsequent iterations, preventing over-approximation.
Exhaustive mode is available for processing all unresolved calls $\mathcal{M}$ (Equation~\ref{eq:missing_calls}) in a single pass when iterative refinement is not required.


\subsection{Security Results}
\label{sec:security_results}
After extracting sources, sinks, and call edges, \toolname runs CodeQL's CWE-specific security queries with the enhanced taint specifications.
We configure CodeQL to emit results in SARIF format~\cite{GitHub_SARIF_Support}, where path-based results include \texttt{threadFlows}: ordered sequences of code locations showing how taint propagates from source to sink~\cite{SARIF}.
We parse these flows and organize alerts by unique sink locations, treating each distinct sink as a separate security finding.
When alerts contain candidate flow summaries, \toolname invokes the \flowSummaryAgentName Agent to validate taint semantics before finalizing results.

\subsection{Flow Summary Agent}
\label{sec:FlowSummaryAgent}

The \flowSummaryAgentName Agent (Figure~\ref{fig:system_arch}~\squaretext{custompurple}{8}) validates \thirdPartyEdgeName flow summaries $E^{\uparrow}_{third}$ (Section~\ref{sec:InterproceduralFlows}) that appear in reported vulnerability paths, determining whether each third-party call propagates or sanitizes taint for the target CWE.

Before validation, \toolname creates a set of all candidate flow summaries across all threadflows for a given CWE.
The agent receives this set along with CWE context, package source code, and access to the same code repositories as the \sourceSinkAgentName Agent: \texttt{node\_modules}, Node.js built-in modules, and the Browser API index.
For each candidate, the agent traces taint flow from call sites,  through external libraries to determine taint semantics.
Appendix~\ref{appendix:flow-summary-agent-walkthrough} demonstrates this analysis for a vulnerability from our validation dataset that requires inspection of \texttt{node\_modules}.

If the agent determines that taint propagates, the call edge remains unchanged and all associated threadflows remain valid alerts.
If taint is sanitized, \toolname marks the edge as ignored, filtering all dependent threadflows from final results.
Flow summaries are saved as JSON documents with metadata supporting manual auditing.




\subsection{Optimizations}
\label{sec:optimizations}

A key part of CodeQL's utility comes from encoding abstract vulnerability patterns as queries that apply universally across packages.
These patterns are pre-compiled and distributed as query packs for fast execution without per-package modification~\cite{Github_QL_PACKS}.
Previous approaches integrating LLMs with CodeQL modify query files directly, embedding extracted sources and sinks into each CWE-specific query~\cite{QLPRO, IRIS}.
These approaches suffer from three limitations:

\begin{description}
    \item[(i)] \textbf{Query recompilation:} Each analysis run requires recompiling modified queries, imposing overhead.
    \item[(ii)] \textbf{Tight coupling:} Specifications become embedded within query text, preventing distribution as reusable query packs or transfer across packages.
    \item[(iii)] \textbf{Single-query scope:} Specifications are injected only into queries that guided their extraction, preventing reuse across different vulnerability types within the same package.
\end{description}

To address these limitations (\textbf{C3}), \toolname leverages CodeQL's external predicate mechanism, which allows predicates to consume data from external files at query evaluation time without recompilation.
We structure taint specifications, sources ($S^{\!\uparrow}$), sinks ($K^{\!\uparrow}$), and call edges ($E^{\!\uparrow}$) as CSV relations that CodeQL imports during analysis.
This decoupled design eliminates recompilation overhead and enables iterative refinement.
Moreover, while source and sink specifications remain most relevant to their guiding CWE, repaired call edges enhance the foundational inter-procedural data flow graph shared by \textit{all} queries, compounding across subsequent analyses.

\section{Evaluation}

This section evaluates \toolname by answering the following research questions.

\begin{enumerate}
    \item[\textbf{RQ1}] \textit{To what extent does \toolname improve CodeQL's ability to discover vulnerabilities?}
    \item[\textbf{RQ2}] \textit{How effectively does \toolname focus agent inference for vulnerability detection?}
    \item[\textbf{RQ3}] \textit{Which \toolname components drive vulnerability detection?}
    \item[\textbf{RQ4}] \textit{Can \toolname find unknown vulnerabilities?}
\end{enumerate}

\subsection{Experimental Setup}

To answer these research questions, we evaluate \toolname on a curated dataset of known npm vulnerabilities to measure recall against ground truth, and on popular open-source packages to assess discovery of previously unknown vulnerabilities.

\subsubsection{Known Vulnerability Dataset} 
\label{sec:known-vuln-dataset}
We use the annotated vulnerability dataset from Brito et al.~\cite{Dataset}, which contains 957 manually verified vulnerabilities in Node.js packages constructed from npm security advisories published through June~2021.
We apply six filters to identify vulnerabilities within \toolname's scope that CodeQL currently misses:

\begin{enumerate}
    \item Remove entries with incomplete metadata or unavailable artifacts (19 entries).
    \item Remove entries whose CWEs are not among CodeQL's 55 supported CWEs (336 entries).
    \item Remove vulnerabilities that CodeQL \codeqlV detects using standard and experimental security queries with and without sanitization (290 entries).
    \item Remove entries whose CWEs lack taint-tracking query support among CodeQL's 37 taint-flow CWEs (49 entries).
    \item Remove entries lacking source/sink location annotations required for ground-truth matching (70 entries).
    \item Remove packages exceeding 1 million tokens (21 entries), a threshold consistent with LLM-based code analysis benchmarks~\cite{ swebench-verified}.
\end{enumerate}

This yields our final dataset of \emph{172} vulnerabilities. 

\myparagraph{Validation Dataset}
We set aside \textit{10} vulnerabilities from our dataset of 172 as a validation set.
We deliberately selected these vulnerabilities to target key failure modes: in three cases, an unresolved call along the source–sink path caused CodeQL to miss the vulnerability (testing the \callgraphAgentName Agent); in two cases, CodeQL's default source/sink configuration was insufficient (testing the \sourceSinkAgentName Agent); the remaining five were sampled randomly for CWE diversity.
All validation entries were constrained to packages under 100k tokens (measured via tiktoken~\cite{Tiktoken_2025}).
This set enabled iterative refinement of agent prompts, hyperparameters, and tool-calling behavior without contaminating evaluation results.

\myparagraph{Testing Dataset}
The remaining \textit{162} vulnerabilities constitute our test set, spanning 17 CWEs across diverse package architectures.

\subsubsection{OSS Dataset}
To evaluate \toolname's ability to discover unknown vulnerabilities, we construct a dataset of popular npm packages from the Libraries.io API~\cite{Libraries_IO_2025}.
We retrieve the top 5,000 packages ranked by the SourceRank algorithm~\cite{Sourceranks_2025} and apply four filters:

\begin{enumerate}
    \item Remove packages with fewer than 1000 dependents or 100 GitHub stars, missing repository URLs, non-JavaScript/TypeScript code, or incomplete metadata (851 entries).
    \item Remove packages exceeding 100k tokens for tractable analysis and manual validation (2,625 entries).
    \item Remove packages without GitHub Security Advisory (GHSA) records~\cite{GitHub_Security_Advisories_2025}, targeting packages with prior vulnerability history as likely candidates (1,368 entries).
    \item Remove packages whose historical vulnerabilities do not match the 37 CWEs that CodeQL's taint-tracking queries support (54 entries).
\end{enumerate}

This yields our final OSS dataset of \emph{102} packages.



\subsubsection{Experimental Design}
\label{sec:experimental-design}
All experiments use CodeQL \codeqlV~\cite{QLVERSION}, a recent stable release.
CodeQL's data flow engine employs conservative budget limits to maintain tractable analysis times.
Two parameters are relevant: \texttt{accessPathLimit}~\cite{CODEQL_ACCESS_PATHS} bounds the depth of chained property accesses the engine tracks, while \texttt{fieldFlowBranchLimit} bounds branching during field flow analysis.
JavaScript queries default these to two, which can prune flows through deeply nested structures.
To verify that these limits do not confound our results, we re-ran the baseline CodeQL configuration on the Known Vulnerability Dataset with both parameters set to 5000 (CodeQL's recommended debugging configuration)~\cite{CODEQL_DEBUGGING}.
This revealed zero additional vulnerabilities for baseline CodeQL, confirming that \toolname's detection improvements stem from source/sink identification and call graph repair rather than circumventing CodeQL's budget constraints.

\myparagraph{Model Choice}
We assign models based on task complexity: the \callgraphAgentName Agent performs structurally constrained call resolution, while the \sourceSinkAgentName and \flowSummaryAgentName Agents require open-ended reasoning across multiple code repositories.
We compared GPT-based (GPT-5, GPT-5-mini) and Gemini-based (Gemini-2.5-Pro, Gemini-2.5-Flash) configurations, pairing higher-capacity models with the \sourceSinkAgentName and \flowSummaryAgentName Agents and lighter-weight models with the \callgraphAgentName Agent.
Table~\ref{table:val-model-combined} shows that GPT-based configurations achieved higher recall; all subsequent experiments use \textbf{GPT-5} for the \sourceSinkAgentName and \flowSummaryAgentName Agents and \textbf{GPT-5-mini} for the \callgraphAgentName Agent.
We run each agent component 3 times, applying union-aggregation for the \sourceSinkAgentName and \callgraphAgentName Agents and majority-vote for the \flowSummaryAgentName Agent, as prior work finds three runs sufficient to smooth LLM variability~\cite{VULN_REPAIR_2025,Ensemble_2025}.


\subsubsection{Evaluation Rulesets}

To isolate the contribution of each \toolname component, we employ an ablation design using seven rulesets that vary across three dimensions: source/sink specifications, call graph edges, and sanitization barriers.
Table~\ref{table:rulesets} defines these rulesets.
$R_1$ represents CodeQL's unmodified baseline configuration, establishing which vulnerabilities require \toolname enhancements.
The hierarchy enables us to determine which enhancements were necessary for detection: if a vulnerability is first detected by $R_2$, enhanced call edges alone were sufficient (with CodeQL's default sources/sinks); if first detected by $R_3$, custom sources/sinks alone were sufficient; if first detected by $R_4$, both were required.
$R_5$ and $R_6$ combine CodeQL's default sources/sinks with \toolname's additions, testing whether the two complement each other.
$R_7$ disables CodeQL's sanitization rules, revealing whether over-aggressive barriers block valid taint flows even after our enhancements.
This design reveals precisely which components drive detection improvements.

\newcommand{\testingNoResult}{27\xspace}
\newcommand{\testingFullMatch}{39\xspace}
\newcommand{\testingThreadMatchTotal}{29\xspace}
\newcommand{\testingThreadMatchActual}{28\xspace}
\newcommand{\testingNoMatchTotal}{67\xspace} 
\newcommand{\testingNoMatchActual}{39\xspace} 
\newcommand{\testingNoMatchFN}{28\xspace} 

\begin{table}[t]
\centering
\footnotesize
\caption{Validation Dataset Results}
\label{table:validation-results}
\setlength{\tabcolsep}{4pt}

\resizebox{\columnwidth}{!}{%
\begin{tabular}{llccccc}
\toprule
\textbf{ID} & \textbf{CWE} & \textbf{Found} & \textbf{Alerts} & \textbf{FP} & \textbf{Method} & \textbf{Ruleset} \\
\midrule
GHSA-3crj-w4f5-gwh4 & CWE-74  & \checkmark & 2  & 0 & TICR & 3 \\
GHSA-5ff8-jcf9-fw62 & CWE-79  & \checkmark & 1  & 0 & TICR & 2 \\
GHSA-8j8c-7jfh-h6hx & CWE-94  & \checkmark & 1  & 0 & EXH  & 2 \\
GHSA-cqjg-whmm-8gv6 & CWE-400 & x           & 11 & 4 & TICR & 4 \\
GHSA-f4hq-453j-p95f & CWE-601 & \checkmark & 1  & 0 & TICR & 2 \\
GHSA-hpr5-wp7c-hh5q\textsuperscript{$\ast$} & CWE-79  & \checkmark & 5  & 0 & TICR & 4 \\
GHSA-jv35-xqg7-f92r & CWE-915 & \checkmark & 5  & 2 & TICR & 2 \\
GHSA-r4m5-47cq-6qg8 & CWE-918 & \checkmark & 1  & 0 & TICR & 2 \\
GHSA-x67x-98x7-wv26 & CWE-78  & \checkmark & 1  & 0 & TICR & 3 \\
GHSA-xqh8-5j36-4556 & CWE-89  & \checkmark & 2  & 0 & TICR & 3 \\
\midrule
\multicolumn{2}{c}{\textbf{Summary}} & \textbf{Found} & \textbf{Alerts} & \textbf{FP} & \textbf{Precision} &  \\
\multicolumn{2}{c}{} & \textbf{9/10} & \textbf{30} & \textbf{6} & \textbf{0.80} &  \\
\bottomrule
\end{tabular}
}

\vspace{2pt}
{\footnotesize\raggedright
\textsuperscript{$\ast$}\,Required \texttt{accessPathLimit}$=$3 (JavaScript default: 2) \par}

\normalsize
\end{table}

\begin{table}[t]
\centering
\footnotesize
\caption{True Positives (TP) and False Negatives (FN) by CWE}
\label{testingRecall}

\resizebox{\columnwidth}{!}{%
\tiny
\renewcommand{\arraystretch}{0.9}
\begin{tabular}{
l@{\hspace{15pt}}
c@{\hspace{20pt}}
c@{\hspace{20pt}}
c@{\hspace{15pt}}
c
}
\toprule
\textbf{CWE} &
\textbf{TP} &
\textbf{FN} &
\textbf{Total} &
\textbf{Recall (\%)} \\
\midrule
CWE-400 & 26 & 16 & 42 & 61.90 \\
CWE-22  & 23 & 5  & 28 & 82.14 \\
CWE-79  & 17 & 11 & 28 & 60.71 \\
CWE-78  & 11 & 3  & 14 & 78.57 \\
CWE-94  & 7  & 4  & 11 & 63.64 \\
CWE-89  & 4  & 4  & 8  & 50.00 \\
CWE-915 & 6  & 2  & 8  & 75.00 \\
CWE-730 & 4  & 1  & 5  & 80.00 \\
CWE-502 & 2  & 1  & 3  & 66.67 \\
CWE-200 & 1  & 2  & 3  & 33.33 \\
CWE-201 & 2  & 1  & 3  & 66.67 \\
CWE-346 & 1  & 2  & 3  & 33.33 \\
CWE-601 & 1  & 1  & 2  & 50.00 \\
CWE-74  & 1  & 0  & 1  & 100.00 \\
CWE-829 & 0  & 1  & 1  & 0.00 \\
CWE-918 & 0  & 1  & 1  & 0.00 \\
CWE-327 & 0  & 1  & 1  & 0.00 \\
\midrule
\textbf{Total} & \textbf{106} & \textbf{56} & \textbf{162} & \textbf{65.43} \\
\bottomrule
\end{tabular}
}
\normalsize
\end{table}

\subsection{RQ1: Vulnerability Recall}
\label{sec:RQ1}

RQ1 measures the extent to which \toolname recovers vulnerabilities that CodeQL currently misses, quantifying recall on a curated dataset of known npm vulnerabilities.

\myparagraph{Validation Dataset} 
\toolname detected 9 of 10 vulnerabilities in the validation dataset, generating 30 alerts with 6 false positives (80\% precision), as shown in Table~\ref{table:validation-results}.
Two independent reviewers examined each alert, marking it as a false positive only when it clearly did not represent any possible vulnerability.

For \texttt{GHSA-hpr5-wp7c-hh5q} (Section~\ref{sec:motivation}), \toolname correctly identified sources, sinks, and call edges, resolving all inter-procedural gaps.
Detection additionally required raising CodeQL's \texttt{accessPathLimit} (Section~\ref{sec:experimental-design}) from 2 to 3 due to deep property chaining in the token rendering pipeline. This was the only case where we had to raise the default budget once the path was reconnected.

For \texttt{GHSA-cqjg-whmm-8gv6}, \toolname established the correct vulnerability path but the \flowSummaryAgentName Agent incorrectly classified a third-party call as sanitizing, causing the missed detection.

\myparagraph{Testing Dataset}
Table~\ref{testingRecall} shows that \toolname produced valid alerts for \testingTP of \testingTotal vulnerabilities (\testingRecall).
We verified results through three categories:

\begin{enumerate}
    \item \textbf{Exact match} (\testingFullMatch): Alert source and sink both matched annotated ground truth. All \testingFullMatch were graded as TPs (100\%).
    \item \textbf{Extended path} (\testingThreadMatchTotal): Both annotated endpoints appeared on the path, but \toolname extended the flow with a new source upstream or new sink downstream. We manually verified that (a) the new endpoint was valid for the vulnerability, and (b) a valid flow path connected the new and annotated endpoints. This confirmed \testingThreadMatchActual TPs (96.5\%).
    \item \textbf{Equivalent flow} (\testingNoMatchTotal): One or both annotated endpoints were absent from the path. We manually verified whether the alert represented the same vulnerability through a different source, sink, or both. This confirmed \testingNoMatchActual TPs (58.2\%).
\end{enumerate}

Two reviewers independently classified each alert in categories 2 and 3.
Uncertainty required joint review until both reached consensus.
Entries producing no alerts (\testingNoResult) were classified as false negatives.
Alerts representing unrelated vulnerabilities were not counted as TPs.
We did not investigate false positives in the testing dataset, as our focus is recall improvement over CodeQL's baseline.
\begin{keyfindingcolorbox}
\textbf{Key Findings:} \toolname detected \testingTP of \testingTotal previously undetectable vulnerabilities \textbf{(\testingRecall recall)}, compared to CodeQL's 0\% baseline.
\end{keyfindingcolorbox}

\subsection{RQ2: Focused Agent Inference}
\label{sec:RQ2}

RQ2 evaluates how effectively \toolname focuses agent inference for vulnerability detection.
We examine TICR for call graph repair (\textbf{C1}) and demand-driven validation for third-party analysis (\textbf{C2}), measuring inference savings on the testing dataset.



\myparagraph{TICR (C1)}
Across the 162 packages in our testing dataset, $|\mathcal{M}|$ totals 94,909 unresolved calls (Section~\ref{sec:callgraph-unresolved-calls}); exhaustively resolving all of them would be prohibitively expensive.
TICR uses taint-reachability to identify the subset lying on potential vulnerability paths, reducing the candidate set to 10,184 calls (an 89.2\% reduction).
Of these, \toolname auto-assigns 4,989 as third-party via custom CodeQL queries, leaving 5,195 calls for agent resolution.
This two-stage filtering reduces agent inference by 94.5\% overall.
On a per-package basis, average agent calls drop from 585.9 to 32.1.
Despite analyzing only 5.5\% of unresolved calls, \toolname achieves 65.43\% recall; exhaustive resolution was necessary for only 1 of 10 validation vulnerabilities.

\myparagraph{Candidate Flow Summaries (C2)}
Third-party calls pose a similar scalability challenge: analyzing every external dependency for taint semantics is infeasible.
\toolname defers this analysis until candidate flow summaries appear in reported vulnerability paths.
This demand-driven strategy reduces third-party edge analysis by 93.2\%, from 6,639 candidates to just 449.
The vast majority of third-party calls never participate in vulnerability paths and are never analyzed.

\begin{keyfindingcolorbox}
\textbf{Key Findings:} By focusing inference on security-relevant calls, \toolname reduces LLM invocations by \textbf{94.5\%} for call graph repair and \textbf{93.2\%} for third-party validation, while achieving 65.43\% recall.
\end{keyfindingcolorbox}

\begin{table}[t]
\centering
\caption{Evaluation Rulesets for Ablation Study}
\label{table:rulesets}

\begin{tabular}{
c@{\hspace{15pt}}
l@{\hspace{25pt}}
c@{\hspace{25pt}}
c
}
\toprule
\textbf{Ruleset} &
\textbf{Sources/Sinks} &
\textbf{Call Graph} &
\textbf{Barriers} \\
\midrule
$R_1$ & Base (CodeQL)          & Base     & Enabled  \\
$R_2$ & Base (CodeQL)          & Enhanced & Enabled  \\
$R_3$ & Custom (\toolname)     & Base     & Enabled  \\
$R_4$ & Custom (\toolname)     & Enhanced & Enabled  \\
$R_5$ & Combined               & Base     & Enabled  \\
$R_6$ & Combined               & Enhanced & Enabled  \\
$R_7$ & Combined               & Enhanced & Disabled \\
\bottomrule
\end{tabular}
\end{table}

\begin{table}[t]
\centering
\footnotesize
\caption{True Positives by Ruleset Group}
\label{table:ruleset_results}
\setlength{\tabcolsep}{2pt}
\resizebox{\columnwidth}{!}{%
\begin{tabular}{lccccccccc}
\toprule
\textbf{Group} & \bm{$R_1$} & \bm{$R_2$} & \bm{$R_3$} & \bm{$R_4$} & \bm{$R_5$} & \bm{$R_6$} & \bm{$R_7$} & \textbf{Total} & \textbf{Percentage (\%)} \\
\midrule
Just source/sink & - & - & 81 & - & 0 & - & - & 81 & 76.42 \\
Needed call graph & - & 3 & - & 15 & - & 3 & 4 & 25 & 23.58 \\
\midrule
\textbf{Total} & \textbf{0} & \textbf{3} & \textbf{81} & \textbf{15} & \textbf{0} & \textbf{3} & \textbf{4} & \textbf{106} & \textbf{100.00} \\
\bottomrule
\end{tabular}
}
\end{table}

\subsection{RQ3: Ablation Study}
\label{sec:RQ3}

RQ3 uses an ablation study to quantify the contribution of each \toolname component and identify which gaps in CodeQL's current capabilities prevented detection.
We report results across all 115 detected vulnerabilities from both datasets: 9 from validation, 106 from testing.


\myparagraph{Component Attribution}
To understand which \toolname components drive vulnerability detection, we evaluated each package using the seven rulesets defined in Table~\ref{table:rulesets}.
This ablation design isolates the contribution of enhanced call graphs ($R_2$), source/sink specifications ($R_3$, $R_5$), and their combination with call graph repair ($R_4$, $R_6$, $R_7$).
We attribute each vulnerability to the first ruleset that successfully detects it.

Table~\ref{table:ruleset_results} presents the testing dataset results; Table~\ref{table:validation-results} presents validation results.
CodeQL's baseline configuration ($R_1$) detected none (0\%) of these 115 vulnerabilities, confirming that each detection requires at least one enhancement provided by \toolname.
Across both datasets, 84 vulnerabilities (73.0\%) were first detected at $R_3$, where CodeQL's existing call graph was sufficient once custom sources and sinks were specified.
The remaining 31 vulnerabilities (27.0\%) required call graph repair, representing a hard detection ceiling: those vulnerabilities fundamentally undetectable without resolving missing inter-procedural edges, regardless of source and sink accuracy.

\myparagraph{Synergy Between Components}
The ablation reveals an important synergy between \toolname's contributions.
Of the 31 vulnerabilities requiring call graph repair, only 8 were detectable with CodeQL's default sources and sinks ($R2$).
The remaining 23 required both custom sources/sinks and call graph repair ($R4$, $R6$, $R7$).
Custom sources and sinks identify package-specific entry points and dangerous operations that CodeQL's defaults miss, revealing new taint paths, but these paths can traverse the same dynamic JavaScript patterns that produce unresolved calls in CodeQL's call graph.
Without call graph repair, the paths remain severed even when both endpoints are correctly specified.

\begin{keyfindingcolorbox}
\textbf{Key Findings:} \toolname closes both semantic and structural gaps in CodeQL. Package-specific sources/sinks enable 73\% of our detections, but the remaining 27\% become visible only after repairing unresolved dynamic calls.
\end{keyfindingcolorbox}

\subsection{RQ4: Unknown Vulnerabilities}
\label{sec:RQ4}
RQ4 evaluates whether \toolname can uncover previously unknown vulnerabilities in popular npm packages.

\myparagraph{Criteria} We consider a vulnerability detected by \toolname novel if: (1) there does not exist a GHSA or CVE that documents the vulnerability for the affected package version, (2) CodeQL's baseline ($R_1$) cannot discover it, and (3) we can create a proof-of-concept (PoC) to demonstrate its exploitability.

\myparagraph{Results}
At the time of archiving, we completed end-to-end validation on \totalOss packages within our OSS Dataset, as PoC-level confirmation and coordinated disclosure introduce substantial manual overhead.
Across this validated subset, \toolname produced 34 alerts.
13 were confirmed as true positives, corresponding to 4 distinct novel vulnerabilities, each affecting a different package.
The remaining 21 alerts were divided into two categories.
(1) \textit{Improper Sanitization Modeling}: Ten contained correctly identified source-to-sink paths but were properly sanitized. Since \toolname delegates sanitizer modeling to CodeQL by design (Section~\ref{sec:sanitizers}), these reflect limitations in CodeQL's existing sanitizer specifications.
(2) \textit{Intended Functionality}: Eleven identified valid source-to-sink paths; however, these corresponded to intended package functionality.
For example, a CLI wrapper package forwards user input to shell execution, which \toolname correctly identifies as a taint flow, but it represents intended functionality rather than a security flaw.
Filtering such cases requires end-to-end reasoning over vulnerability paths with knowledge of each package's intended functionality. This analysis lies outside the scope of taint specification extraction.

The 4 confirmed vulnerabilities are distributed across four distinct packages and span three CWE classes: command injection (CWE-78), path traversal (CWE-22), and prototype pollution (CWE-915).
To avoid premature disclosure, we omit package identifiers and exploit details.
Overall, these results indicate that \toolname can uncover previously undocumented, exploitable vulnerabilities in widely used packages that are missed by a CodeQL baseline configuration.




\begin{keyfindingcolorbox}
\textbf{Key Findings:} \toolname discovered \ossResults previously unknown vulnerabilities in 4 popular npm packages. These findings demonstrate that \toolname's LLM-augmented taint specification extraction enables discovery of real-world vulnerabilities that CodeQL cannot detect.
\end{keyfindingcolorbox}

\subsection{Threats to Validity}

\myparagraph{Internal Validity}
LLMs are inherently non-deterministic, which affects reliability of results.
We mitigate this through multi-run aggregation, taking the union of results across three executions per agent.

Alert classification relied on heuristic matching against Brito et al.'s~\cite{Dataset} annotated source/sink pairs.
Exact matches were automatic true positives, while extended or equivalent paths required verification from two independent reviewers.
This process may miss valid alerts that differ structurally from annotations and inherit any labeling errors present in the original dataset.

\ThirdPartyEdgeName flow summaries model taint as returning through invocation sites, missing cases where third-party libraries pass taint to other code locations through callbacks.
Modeling callback-mediated propagation would require full review of each third-party unresolved call, substantially increasing inference cost.

TICR's taint-reachability analysis can miss security-relevant calls when intra-procedural gaps sever paths before reaching unresolved calls.
Extended flow semantics (Figure~\ref{fig:extended_flow_rules}) partially address this limitation, and exhaustive mode provides a fallback for comprehensive coverage.

\myparagraph{External Validity}
We evaluate only JavaScript/npm packages.
Extension to other languages requires reimplementation of the CodeQL integration layer, though the agent architecture and TICR algorithm are language-agnostic in principle.
Our 1-million-token limit excludes 21 large packages from the dataset, and we cover only the 37 CWEs supported by CodeQL's taint-tracking queries.

The Brito et al.\ dataset contains vulnerabilities disclosed through June 2021.
JavaScript development practices and npm package structures evolve, which may limit generalizability to contemporary codebases.
Similarly, our OSS dataset filters for packages with prior GHSA records, which may bias toward packages with known vulnerability history.

Several limitations are explicitly out of scope: intra-procedural data flow gaps arising from JavaScript's dynamic features and CodeQL's sanitizer over-approximation.
These represent fundamental challenges in JavaScript static analysis that \toolname does not address.

\section{Related Work}

\myparagraph{Learning-Based Call Graph Repair} 
CALLME~\cite{wang2025callme} uses an LLM to resolve only dynamic property accesses in JavaScript from a dataset of unresolved calls. 
CALLME requires a separate LLM query for each call-site to candidate-function pair, limiting scalability for entire call graph construction.
GRAPHIA~\cite{callmemaybe} uses graph neural networks to predict unresolved call edges within JavaScript projects. 
This approach requires training a separate model for each package, using that package's statically-resolved edges, whereas \toolname leverages pre-trained LLMs that generalize across packages without package-specific training.
Additionally, both CALLME and GRAPHIA operate without integrating call graph repair within SAST tools for vulnerability detection. 
Moreover, neither approach addresses third-party dependencies, resolving calls only to functions defined within the analyzed project.


\myparagraph{LLM SAST Integration} 
IRIS~\cite{IRIS} and QLPro~\cite{QLPRO} use LLMs to infer sources and sinks and generate CodeQL queries for Java programs. 
However, both rely entirely on the static analyzer's existing call graph to connect identified endpoints. Both process APIs in batches constrained by LLM context windows, limiting analysis of large codebases and third-party dependencies. \toolname overcomes both limitations: it repairs broken inter-procedural edges preventing taint propagation, and its agents explore code on-demand, encompassing dependency implementations.


\myparagraph{JavaScript Taint Specification Extraction}
TASER extracts taint specifications for JavaScript libraries via dynamic taint tracking driven by existing test suites of the library and its npm clients~\cite{TASER}. 
As a dynamic technique, it only observes executed behavior, so the completeness of extracted specifications depends on the availability and coverage of those tests.
InspectJS infers sink specifications from static data-flow information in CodeQL and refines the resulting predictions using code similarity and user feedback, ultimately to improve manually maintained models~\cite{INSPECTJS}.
In contrast, \toolname produces taint specifications (sources, sinks, call edges, and flow summaries), without relying on existing test coverage or user feedback.

\myparagraph{Taint-Guided Call Graph Construction} 
TAJ (Taint Analysis for Java)~\cite{TAJ} uses a priority-driven strategy that ranks methods by graph distance from taint sources rather than using taint-flows to determine reachability. 
Seneca~\cite{SENECA} uses taint state to guide call resolution at Java deserialization points, conservatively expanding targets to any serializable type that could appear at runtime.
TICR differs from prior works in three key respects: (1) it uses taint reachability to identify \emph{which} unresolved calls require repair, rather than determining \emph{how} to resolve known call targets; (2) it performs bidirectional analysis, identifying both unresolved calls reachable from sources and those whose return values can reach sinks (Figures~\ref{fig:Source-Break}~and~\ref{fig:Break-Sink}), rather than unidirectional tracking from sources; and (3) it targets JavaScript, where dynamic features create unique taint-tracking challenges requiring custom propagation semantics (Figure~\ref{fig:extended_flow_rules}).

\section{Conclusion}
JavaScript is particularly difficult to statically analyze for vulnerabilities.
This paper proposed \toolname, a novel approach for combining traditional SAST tools with LLMs, using the SAST tool to drive the analysis, and only deferring to an LLM where necessary.
Our design breaks the analysis problem into three key parts.
A Source/Sink Agent identifies taint sources and sinks unique to a target JavaScript package.
A \callgraphAgentName Agent resolves call edges that could not be resolved by the SAST tool, intelligently doing so only for edges that potentially lead to a vulnerability.
A Flow Summary Agent refines candidate vulnerable flows, performing in-depth analysis of third-party packages to filter out non-vulnerable cases.
We implemented a prototype of \toolname for CodeQL and evaluated it on \testingTotal known vulnerabilities that CodeQL alone could not detect.
By enhancing CodeQL with \toolname, the analysis found \testingTP (\testingRecall) of the previously undetectable vulnerabilities.
We further deployed \toolname on open-source packages and identified \ossResultsWord previously unknown vulnerabilities.
In doing so, we demonstrate how LLMs can practically enhance existing SAST tools by combining the strengths of symbolic reasoning and semantic understanding.




\bibliographystyle{IEEEtran}
\bibliography{ref}

@article{static&dynamicCOMP,
  title={Is JavaScript Call Graph Extraction Solved Yet? A Comparative Study of Static and Dynamic Tools},
  author={G{\'a}bor Antal and P{\'e}ter Hegedűs and Zolt{\'a}n Herczeg and G{\'a}bor L{\'o}ki and Rudolf Ferenc},
  journal={IEEE Access},
  year={2023},
  volume={11},
  pages={25266-25284},
  url={https://api.semanticscholar.org/CorpusID:257480090}
}

@article{Dataset,
   title={Study of JavaScript Static Analysis Tools for Vulnerability Detection in Node.js Packages},
   volume={72},
   ISSN={1558-1721},
   url={http://dx.doi.org/10.1109/TR.2023.3286301},
   DOI={10.1109/tr.2023.3286301},
   number={4},
   journal={IEEE Transactions on Reliability},
   publisher={Institute of Electrical and Electronics Engineers (IEEE)},
   author={Brito, Tiago and Ferreira, Mafalda and Monteiro, Miguel and Lopes, Pedro and Barros, Miguel and Santos, José Fragoso and Santos, Nuno},
   year={2023},
   month=dec, pages={1324–1339} }

@inproceedings{Park_2021, series={ESEC/FSE ’21},
   title={Accelerating JavaScript static analysis via dynamic shortcuts},
   url={http://dx.doi.org/10.1145/3468264.3468556},
   DOI={10.1145/3468264.3468556},
   booktitle={Proceedings of the 29th ACM Joint Meeting on European Software Engineering Conference and Symposium on the Foundations of Software Engineering},
   publisher={ACM},
   author={Park, Joonyoung and Park, Jihyeok and Youn, Dongjun and Ryu, Sukyoung},
   year={2021},
   month=aug, collection={ESEC/FSE ’21} }

@inproceedings{STRINGDOMAINS,
author = {Amadini, Roberto and Jordan, Alexander and Gange, Graeme and Gauthier, Fran\c{c}ois and Schachte, Peter and S\O{}ndergaard, Harald and Stuckey, Peter J. and Zhang, Chenyi},
title = {Combining String Abstract Domains for JavaScript Analysis: An Evaluation},
year = {2017},
isbn = {9783662545768},
publisher = {Springer-Verlag},
address = {Berlin, Heidelberg},
url = {https://doi.org/10.1007/978-3-662-54577-5_3},
doi = {10.1007/978-3-662-54577-5_3},
booktitle = {Proceedings, Part I, of the 23rd International Conference on Tools and Algorithms for the Construction and Analysis of Systems - Volume 10205},
pages = {41–57},
numpages = {17}
}

@misc{Stack_Overflow_Developer_Survey_2025,
  author       = {{Stack Overflow}},
  title        = {2025 {Stack Overflow} Developer Survey},
  howpublished = {https://survey.stackoverflow.co/2025/technology},
  year         = {2025}
}

@inproceedings{INSPECTJS,
author = {Dutta, Saikat and Garbervetsky, Diego and Lahiri, Shuvendu K. and Sch\"{a}fer, Max},
title = {InspectJS: leveraging code similarity and user-feedback for effective taint specification inference for JavaScript},
year = {2022},
isbn = {9781450392266},
publisher = {Association for Computing Machinery},
address = {New York, NY, USA},
url = {https://doi.org/10.1145/3510457.3513048},
doi = {10.1145/3510457.3513048},
booktitle = {Proceedings of the 44th International Conference on Software Engineering: Software Engineering in Practice},
pages = {165–174},
numpages = {10},
location = {Pittsburgh, Pennsylvania},
series = {ICSE-SEIP '22}
}

@inproceedings{TASER,
author = {Staicu, Cristian-Alexandru and Torp, Martin Toldam and Sch\"{a}fer, Max and M\o{}ller, Anders and Pradel, Michael},
title = {Extracting taint specifications for JavaScript libraries},
year = {2020},
isbn = {9781450371216},
publisher = {Association for Computing Machinery},
address = {New York, NY, USA},
url = {https://doi.org/10.1145/3377811.3380390},
doi = {10.1145/3377811.3380390},
booktitle = {Proceedings of the ACM/IEEE 42nd International Conference on Software Engineering},
pages = {198–209},
numpages = {12},
location = {Seoul, South Korea},
series = {ICSE '20}
}

@misc{akhavani2025opensourceopenthreats,
      title={Open Source, Open Threats? Investigating Security Challenges in Open-Source Software}, 
      author={Seyed Ali Akhavani and Behzad Ousat and Amin Kharraz},
      year={2025},
      eprint={2506.12995},
      archivePrefix={arXiv},
      primaryClass={cs.CR},
      note={\url{https://arxiv.org/abs/2506.12995}}, 
}

@misc{SAST-GENIUS,
      title={LLM-Driven SAST-Genius: A Hybrid Static Analysis Framework for Comprehensive and Actionable Security}, 
      author={Vaibhav Agrawal and Kiarash Ahi},
      year={2025},
      eprint={2509.15433},
      archivePrefix={arXiv},
      primaryClass={cs.CR},
      note={\url{https://arxiv.org/abs/2509.15433}}, 
}

@misc{CPGLLM,
      title={Utilizing Precise and Complete Code Context to Guide LLM in Automatic False Positive Mitigation}, 
      author={Jinbao Chen and Hongjing Xiang and Zuohong Zhao and Luhao Li and Yu Zhang and Boyao Ding and Qingwei Li and Songyuan Xiong},
      year={2025},
      eprint={2411.03079},
      archivePrefix={arXiv},
      primaryClass={cs.SE},
      note={\url{https://arxiv.org/abs/2411.03079}}, 
}

@misc{IRIS,
      title={IRIS: LLM-Assisted Static Analysis for Detecting Security Vulnerabilities}, 
      author={Ziyang Li and Saikat Dutta and Mayur Naik},
      year={2025},
      eprint={2405.17238},
      archivePrefix={arXiv},
      primaryClass={cs.CR},
      note={\url{https://arxiv.org/abs/2405.17238}}, 
}

@misc{QLPRO,
      title={QLPro: Automated Code Vulnerability Discovery via LLM and Static Code Analysis Integration}, 
      author={Junze Hu and Xiangyu Jin and Yizhe Zeng and Yuling Liu and Yunpeng Li and Dan Du and Kaiyu Xie and Hongsong Zhu},
      year={2025},
      eprint={2506.23644},
      archivePrefix={arXiv},
      primaryClass={cs.SE},
      note={\url{https://arxiv.org/abs/2506.23644}}, 
}

@misc{CodeQL_Node_Modules, title={CodeQL Call-Graph Construction for JavaScript}, note={\url{https://codeql.github.com/docs/codeql-language-guides/codeql-library-for-javascript/}}, journal={GitHub Docs}, author={GitHub}, year={2025}}

@misc{CodeQL_Nodes, title={About data flow analysis}, note={\url{https://codeql.github.com/docs/writing-codeql-queries/about-data-flow-analysis/}}, journal={GitHub Docs}, author={GitHub}, year={2025}}

@misc{CodeQL_Get_A_CALLEE_2025, title={Member predicate dataflow::invokenode}, note ={\url{https://codeql.github.com/codeql-standard-libraries/javascript/semmle/javascript/dataflow/Nodes.qll/predicate.Nodes$InvokeNode$getACallee.0.html}}, journal={getACallee}, author={GitHub}, year={2025}}

@misc{CodeQL_Main_Docs,
  title   = {CodeQL},
  note    = {\url{https://codeql.github.com/}},
  journal = {GitHub Docs},
  author  = {GitHub},
  year    = {2025}
}

@misc{GitHub_SARIF_Support,
  title   = {SARIF support for code scanning},
  author  = {GitHub},
  journal = {GitHub Docs},
  year    = {2026},
  note    = {\url{https://docs.github.com/en/code-security/code-scanning/integrating-with-code-scanning/sarif-support-for-code-scanning}},
}

@misc{SARIF,
  title   = {Static Analysis Results Interchange Format (SARIF) Version 2.1.0: Committee Specification 01},
  author  = {Michael C. Fanning and Laurence J. Golding},
  journal = {OASIS Committee Specification 01},
  year    = {2019},
  note    = {\url{https://docs.oasis-open.org/sarif/sarif/v2.1.0/cs01/sarif-v2.1.0-cs01.pdf}},
}

@inproceedings{
wang2025callme,
title={{CALLME}: Call Graph Augmentation with Large Language Models for Javascript},
author={Michael Wang and Kexin Pei and Armando Solar-Lezama},
booktitle={Second Conference on Language Modeling},
year={2025},
note={https://openreview.net/forum?id=xZi2rMUcAO}
}

@article{SAST_POWER,
author = {Cui, Lei and Cui, Jiancong and Hao, Zhiyu and Li, Lun and Ding, Zhenquan and Liu, Yongji},
title = {An empirical study of vulnerability discovery methods over the past ten years},
year = {2022},
issue_date = {Sep 2022},
publisher = {Elsevier Advanced Technology Publications},
address = {GBR},
volume = {120},
number = {C},
issn = {0167-4048},
url = {https://doi.org/10.1016/j.cose.2022.102817},
doi = {10.1016/j.cose.2022.102817},
journal = {Comput. Secur.},
month = sep,
numpages = {13}
}

@inproceedings{MODULARCALLGRAPH,
author = {Nielsen, Benjamin Barslev and Torp, Martin Toldam and M\o{}ller, Anders},
title = {Modular call graph construction for security scanning of Node.js applications},
year = {2021},
isbn = {9781450384599},
publisher = {Association for Computing Machinery},
address = {New York, NY, USA},
url = {https://doi.org/10.1145/3460319.3464836},
doi = {10.1145/3460319.3464836},
booktitle = {Proceedings of the 30th ACM SIGSOFT International Symposium on Software Testing and Analysis},
pages = {29–41},
numpages = {13},
location = {Virtual, Denmark},
series = {ISSTA 2021}
}

@misc{callmemaybe,
      title={Call Me Maybe: Enhancing JavaScript Call Graph Construction using Graph Neural Networks}, 
      author={Masudul Hasan Masud Bhuiyan and Gianluca De Stefano and Giancarlo Pellegrino and Cristian-Alexandru Staicu},
      year={2025},
      eprint={2506.18191},
      archivePrefix={arXiv},
      primaryClass={cs.SE},
      note={\url{https://arxiv.org/abs/2506.18191}}, 
}

@misc{CODEQL_ACCESS_PATHS,
  howpublished={{\url{https://codeql.github.com/docs/codeql-language-guides/data-flow-cheat-sheet-for-javascript/#access-paths}}},
  title = {{CodeQL JavaScript data flow cheat sheet: Access paths}},
  key={{CodeQL JavaScript data flow cheat sheet: Access paths}}
}

@article{Survey_of_Dynamic_Analysis,
author = {Andreasen, Esben and Gong, Liang and M\o{}ller, Anders and Pradel, Michael and Selakovic, Marija and Sen, Koushik and Staicu, Cristian-Alexandru},
title = {A Survey of Dynamic Analysis and Test Generation for JavaScript},
year = {2017},
issue_date = {September 2018},
publisher = {Association for Computing Machinery},
address = {New York, NY, USA},
volume = {50},
number = {5},
issn = {0360-0300},
url = {https://doi.org/10.1145/3106739},
doi = {10.1145/3106739},
journal = {ACM Comput. Surv.},
month = sep,
articleno = {66},
numpages = {36}
}

@misc{CVEGENIE,
      title={From CVE Entries to Verifiable Exploits: An Automated Multi-Agent Framework for Reproducing CVEs}, 
      author={Saad Ullah and Praneeth Balasubramanian and Wenbo Guo and Amanda Burnett and Hammond Pearce and Christopher Kruegel and Giovanni Vigna and Gianluca Stringhini},
      year={2025},
      eprint={2509.01835},
      archivePrefix={arXiv},
      primaryClass={cs.CR},
      note={\url{https://arxiv.org/abs/2509.01835}}, 
}

@misc{LLMSTABILITY,
      title={Non-Determinism of "Deterministic" LLM Settings}, 
      author={Berk Atil and Sarp Aykent and Alexa Chittams and Lisheng Fu and Rebecca J. Passonneau and Evan Radcliffe and Guru Rajan Rajagopal and Adam Sloan and Tomasz Tudrej and Ferhan Ture and Zhe Wu and Lixinyu Xu and Breck Baldwin},
      year={2025},
      eprint={2408.04667},
      archivePrefix={arXiv},
      primaryClass={cs.CL},
      note={\url{https://arxiv.org/abs/2408.04667}}, 
}

@inproceedings {ODGen,
author = {Song Li and Mingqing Kang and Jianwei Hou and Yinzhi Cao},
title = {Mining Node.js Vulnerabilities via Object Dependence Graph and Query},
booktitle = {31st USENIX Security Symposium (USENIX Security 22)},
year = {2022},
isbn = {978-1-939133-31-1},
address = {Boston, MA},
pages = {143--160},
url = {https://www.usenix.org/conference/usenixsecurity22/presentation/li-song},
publisher = {USENIX Association},
month = aug
}

@misc{MITRE_CWE, title={Common weakness enumeration}, note={\url{https://cwe.mitre.org/documents/cwe_usage/guidance.html}}, journal={CWE}, author={MITRE}}

@misc{Github_QL_PACKS, title={Creating and working with CodeQL packs}, note={\url{https://docs.github.com/en/code-security/codeql-cli/using-the-advanced-functionality-of-the-codeql-cli/creating-and-working-with-codeql-packs}}, journal={GitHub Docs}, author={GitHub}, year={2025}}

@misc{aggregation_LLMs,
      title={Self-Consistency Improves Chain of Thought Reasoning in Language Models}, 
      author={Xuezhi Wang and Jason Wei and Dale Schuurmans and Quoc Le and Ed Chi and Sharan Narang and Aakanksha Chowdhery and Denny Zhou},
      year={2023},
      eprint={2203.11171},
      archivePrefix={arXiv},
      primaryClass={cs.CL},
      note={\url{https://arxiv.org/abs/2203.11171}}, 
}

@article{piskachev,
  title={{Adapting Taint Analyses for Detecting Security Vulnerabilities}},
  author={Goran, Piskachev},
  year={2023},
  url={https://www.bodden.de/pubs/thesis-piskachev.pdf}
}

@misc{mrk_js,
  author       = {Alex Bates},
  title        = {mrk: Tiny Extensible Markdown Parser/Renderer (single-file JavaScript library)},
  year         = {2025},
  note          = {\url{https://github.com/bates64/mrk}},
  note         = {Accessed: 2025-11-13}
}

@software{markdown-it-katex,
  title = {Waylonflinn/Markdown-It-Katex},
  author = {Flinn, Waylon},
  date = {2025-10-05T05:27:35Z},
  origdate = {2016-03-11T13:26:47Z},
  url = {https://github.com/waylonflinn/markdown-it-katex},
  urldate = {2025-12-13}
}

@misc{GitHub_Security_Advisories_2025,
  title        = {GitHub Security Advisories},
  note          = {\url{https://github.com/advisories}},
  journal      = {GitHub Advisories},
  author       = {GitHub},
  year         = {2025}
}

@misc{Libraries_IO_2025,
  title   = {Libraries.io},
  note     = {\url{https://libraries.io/}},
  journal = {Libraries.io},
  author  = {Libraries.io},
  year    = {2025}
}

@misc{Sourceranks_2025,
  title   = {Sourceranks},
  note     = {\url{https://github.com/nice-registry/sourceranks}},
  journal = {Sourceranks Repository},
  author  = {Nice Registry},
  year    = {2025}
}

@misc{Tiktoken_2025,
  title   = {tiktoken},
  note     = {\url{https://github.com/openai/tiktoken}},
  journal = {tiktoken Repository},
  author  = {OpenAI},
  year    = {2025}
}

@misc{Ensemble_2025,
      title={A Simple Ensemble Strategy for LLM Inference: Towards More Stable Text Classification}, 
      author={Junichiro Niimi},
      year={2025},
      eprint={2504.18884},
      archivePrefix={arXiv},
      primaryClass={cs.CL},
      note={\url{https://arxiv.org/abs/2504.18884}}, 
}

@misc{VULN_REPAIR_2025,
      title={Identifying Helpful Context for LLM-based Vulnerability Repair: A Preliminary Study}, 
      author={Gábor Antal and Bence Bogenfürst and Rudolf Ferenc and Péter Hegedűs},
      year={2025},
      eprint={2506.11561},
      archivePrefix={arXiv},
      primaryClass={cs.SE},
      note={\url{https://arxiv.org/abs/2506.11561}}, 
}

@inproceedings{jsstatic1,
author = {Kashyap, Vineeth and Dewey, Kyle and Kuefner, Ethan A. and Wagner, John and Gibbons, Kevin and Sarracino, John and Wiedermann, Ben and Hardekopf, Ben},
title = {JSAI: a static analysis platform for JavaScript},
year = {2014},
isbn = {9781450330565},
publisher = {Association for Computing Machinery},
address = {New York, NY, USA},
url = {https://doi.org/10.1145/2635868.2635904},
doi = {10.1145/2635868.2635904},
booktitle = {Proceedings of the 22nd ACM SIGSOFT International Symposium on Foundations of Software Engineering},
pages = {121–132},
numpages = {12},
location = {Hong Kong, China},
series = {FSE 2014}
}

@inproceedings{jsstatic2,
author = {Madsen, Magnus and Livshits, Benjamin and Fanning, Michael},
title = {Practical static analysis of JavaScript applications in the presence of frameworks and libraries},
year = {2013},
isbn = {9781450322379},
publisher = {Association for Computing Machinery},
address = {New York, NY, USA},
url = {https://doi.org/10.1145/2491411.2491417},
doi = {10.1145/2491411.2491417},
booktitle = {Proceedings of the 2013 9th Joint Meeting on Foundations of Software Engineering},
pages = {499–509},
numpages = {11},
location = {Saint Petersburg, Russia},
series = {ESEC/FSE 2013}
}

@article{jsstatic3,
author = {Madsen, Magnus and Tip, Frank and Lhot\'{a}k, Ond\v{r}ej},
title = {Static analysis of event-driven Node.js JavaScript applications},
year = {2015},
issue_date = {October 2015},
publisher = {Association for Computing Machinery},
address = {New York, NY, USA},
volume = {50},
number = {10},
issn = {0362-1340},
url = {https://doi.org/10.1145/2858965.2814272},
doi = {10.1145/2858965.2814272},
journal = {SIGPLAN Not.},
month = oct,
pages = {505–519},
numpages = {15}
}

@inproceedings{CORRTRACKING,
author = {Sridharan, Manu and Dolby, Julian and Chandra, Satish and Sch\"{a}fer, Max and Tip, Frank},
title = {Correlation tracking for points-to analysis of javascript},
year = {2012},
isbn = {9783642310560},
publisher = {Springer-Verlag},
address = {Berlin, Heidelberg},
url = {https://doi.org/10.1007/978-3-642-31057-7_20},
doi = {10.1007/978-3-642-31057-7_20},
booktitle = {Proceedings of the 26th European Conference on Object-Oriented Programming},
pages = {435–458},
numpages = {24},
location = {Beijing, China},
series = {ECOOP'12}
}

@InProceedings{VALUEPARITION,
  author =	{Nielsen, Benjamin Barslev and M{\o}ller, Anders},
  title =	{{Value Partitioning: A Lightweight Approach to Relational Static Analysis for JavaScript}},
  booktitle =	{34th European Conference on Object-Oriented Programming (ECOOP 2020)},
  pages =	{16:1--16:28},
  series =	{Leibniz International Proceedings in Informatics (LIPIcs)},
  ISBN =	{978-3-95977-154-2},
  ISSN =	{1868-8969},
  year =	{2020},
  volume =	{166},
  editor =	{Hirschfeld, Robert and Pape, Tobias},
  publisher =	{Schloss Dagstuhl -- Leibniz-Zentrum f{\"u}r Informatik},
  address =	{Dagstuhl, Germany},
  URL =		{https://drops.dagstuhl.de/entities/document/10.4230/LIPIcs.ECOOP.2020.16},
  URN =		{urn:nbn:de:0030-drops-131731},
  doi =		{10.4230/LIPIcs.ECOOP.2020.16}
}

@inproceedings{richardsAnalysisDynamicBehavior,
author = {Richards, Gregor and Lebresne, Sylvain and Burg, Brian and Vitek, Jan},
title = {An analysis of the dynamic behavior of JavaScript programs},
year = {2010},
isbn = {9781450300193},
publisher = {Association for Computing Machinery},
address = {New York, NY, USA},
url = {https://doi.org/10.1145/1806596.1806598},
doi = {10.1145/1806596.1806598},
booktitle = {Proceedings of the 31st ACM SIGPLAN Conference on Programming Language Design and Implementation},
pages = {1–12},
numpages = {12},
location = {Toronto, Ontario, Canada},
series = {PLDI '10}
}

@misc{LLM_CONFIDENCE,
      title={Can LLMs Express Their Uncertainty? An Empirical Evaluation of Confidence Elicitation in LLMs}, 
      author={Miao Xiong and Zhiyuan Hu and Xinyang Lu and Yifei Li and Jie Fu and Junxian He and Bryan Hooi},
      year={2024},
      eprint={2306.13063},
      archivePrefix={arXiv},
      primaryClass={cs.CL},
      note={\url{https://arxiv.org/abs/2306.13063}}, 
}

@inbook{almashfiStaticTaintAnalysis2021,
author = {Almashfi, Nabil and Lu, Lunjin},
year = {2021},
month = {03},
pages = {155-167},
title = {Static Taint Analysis for JavaScript Programs},
isbn = {978-3-030-71471-0},
doi = {10.1007/978-3-030-71472-7_13}
}

@inproceedings{ANDROMEDA,
author = {Tripp, Omer and Pistoia, Marco and Cousot, Patrick and Cousot, Radhia and Guarnieri, Salvatore},
title = {ANDROMEDA: accurate and scalable security analysis of web applications},
year = {2013},
isbn = {9783642370564},
publisher = {Springer-Verlag},
address = {Berlin, Heidelberg},
url = {https://doi.org/10.1007/978-3-642-37057-1_15},
doi = {10.1007/978-3-642-37057-1_15},
booktitle = {Proceedings of the 16th International Conference on Fundamental Approaches to Software Engineering},
pages = {210–225},
numpages = {16},
location = {Rome, Italy},
series = {FASE'13}
}

@inproceedings{SAVINGTHEWORLD,
author = {Guarnieri, Salvatore and Pistoia, Marco and Tripp, Omer and Dolby, Julian and Teilhet, Stephen and Berg, Ryan},
title = {Saving the world wide web from vulnerable JavaScript},
year = {2011},
isbn = {9781450305624},
publisher = {Association for Computing Machinery},
address = {New York, NY, USA},
url = {https://doi.org/10.1145/2001420.2001442},
doi = {10.1145/2001420.2001442},
booktitle = {Proceedings of the 2011 International Symposium on Software Testing and Analysis},
pages = {177–187},
numpages = {11},
location = {Toronto, Ontario, Canada},
series = {ISSTA '11}
}

@article{PropRules1,
  title={One Engine To Serve 'em All: Inferring Taint Rules Without Architectural Semantics},
  author={Zheng Leong Chua and Yanhao Wang and Teodora Baluta and P. Saxena and Zhenkai Liang and Purui Su},
  journal={Proceedings 2019 Network and Distributed System Security Symposium},
  year={2019},
  url={https://api.semanticscholar.org/CorpusID:96446274}
}

@article{Sanitizers,
  title={A Static Detection Method for SQL Injection Vulnerability Based on Program Transformation},
  author={Ye Yuan and Yuliang Lu and Kailong Zhu and Hui Huang and Lu Yu and Jiazhen Zhao},
  journal={Applied Sciences},
  year={2023},
  url={https://api.semanticscholar.org/CorpusID:264580534}
}

@manual{NodeBuiltinDocs,
  title        = {Modules: CommonJS modules},
  subtitle     = {Built-in modules},
  organization = {Node.js Foundation},
  year         = {2025},
  note         = {Node.js v25.2.1 Documentation},
  url          = {https://nodejs.org/api/modules.html#built-in-modules},
  urldate      = {2025-11-28}
}

@inproceedings{VV8,
author = {Jueckstock, Jordan and Kapravelos, Alexandros},
title = {VisibleV8: In-browser Monitoring of JavaScript in the Wild},
year = {2019},
isbn = {9781450369480},
publisher = {Association for Computing Machinery},
address = {New York, NY, USA},
url = {https://doi.org/10.1145/3355369.3355599},
doi = {10.1145/3355369.3355599},
booktitle = {Proceedings of the Internet Measurement Conference},
pages = {393–405},
numpages = {13},
location = {Amsterdam, Netherlands},
series = {IMC '19}
}

@manual{QLVERSION,
  title        = {CodeQL 2.21.2 Changelog},
  organization = {GitHub},
  year         = {2025},
  note         = {CodeQL CLI 2.21.2 release notes},
  url          = {https://codeql.github.com/docs/codeql-overview/codeql-changelog/codeql-cli-2.21.2/},
  urldate      = {2025-11-29}
}

@inproceedings{NPMSTATS,
author = {Pinckney, Donald and Cassano, Federico and Guha, Arjun and Bell, Jonathan},
title = {npm-follower: A Complete Dataset Tracking the NPM Ecosystem},
year = {2023},
isbn = {9798400703270},
publisher = {Association for Computing Machinery},
address = {New York, NY, USA},
url = {https://doi.org/10.1145/3611643.3613094},
doi = {10.1145/3611643.3613094},
pages = {2132–2136},
numpages = {5},
location = {San Francisco, CA, USA},
series = {ESEC/FSE 2023}
}

@inproceedings{JavaCallgraph1,
author = {Bruce, Bobby R. and Zhang, Tianyi and Arora, Jaspreet and Xu, Guoqing Harry and Kim, Miryung},
title = {JShrink: in-depth investigation into debloating modern Java applications},
year = {2020},
isbn = {9781450370431},
publisher = {Association for Computing Machinery},
address = {New York, NY, USA},
url = {https://doi.org/10.1145/3368089.3409738},
doi = {10.1145/3368089.3409738},
pages = {135–146},
numpages = {12},
location = {Virtual Event, USA},
series = {ESEC/FSE 2020}
}

@inproceedings{JavaCallgraph2,
author = {Sui, Li and Dietrich, Jens and Tahir, Amjed and Fourtounis, George},
title = {On the recall of static call graph construction in practice},
year = {2020},
isbn = {9781450371216},
publisher = {Association for Computing Machinery},
address = {New York, NY, USA},
url = {https://doi.org/10.1145/3377811.3380441},
doi = {10.1145/3377811.3380441},
booktitle = {Proceedings of the ACM/IEEE 42nd International Conference on Software Engineering},
pages = {1049–1060},
numpages = {12},
location = {Seoul, South Korea},
series = {ICSE '20}
}

@inproceedings{PythonCallGraph,
author = {Salis, Vitalis and Sotiropoulos, Thodoris and Louridas, Panos and Spinellis, Diomidis and Mitropoulos, Dimitris},
title = {PyCG: Practical Call Graph Generation in Python},
year = {2021},
isbn = {9781450390859},
publisher = {IEEE Press},
url = {https://doi.org/10.1109/ICSE43902.2021.00146},
doi = {10.1109/ICSE43902.2021.00146},
booktitle = {Proceedings of the 43rd International Conference on Software Engineering},
pages = {1646–1657},
numpages = {12},
location = {Madrid, Spain},
series = {ICSE '21}
}

@article{AverageNPMDeps,
  author       = {Markus Zimmermann and
                  Cristian{-}Alexandru Staicu and
                  Cam Tenny and
                  Michael Pradel},
  title        = {Small World with High Risks: {A} Study of Security Threats in the
                  npm Ecosystem},
  journal      = {CoRR},
  volume       = {abs/1902.09217},
  year         = {2019},
  url          = {http://arxiv.org/abs/1902.09217},
  eprinttype    = {arXiv},
  eprint       = {1902.09217},
  bibsource    = {dblp computer science bibliography, https://dblp.org}
}

@software{VisibleV8Release,
  title = {Wspr-Ncsu/Visiblev8},
  date = {2025-11-24T07:59:55Z},
  origdate = {2019-10-21T17:51:07Z},
  url = {https://github.com/wspr-ncsu/visiblev8/releases/},
  urldate = {2025-12-05},
  organization = {wspr-ncsu}
}

@online{SemGrep_DepIgnore,
  title = {Ignore Files, Folders, and Code | {{Semgrep}}},
  date = {2025-04-04},
  url = {https://semgrep.dev/docs/ignoring-files-folders-code},
  urldate = {2025-12-05},
  langid = {english}
}

@online{SonarQube_DepIgnore,
  title = {{{JavaScript}}/{{TypeScript}}/{{CSS}} | {{SonarQube Server}} 9.9 | {{Sonar Documentation}}},
  date = {2025-08-05},
  url = {https://docs.sonarsource.com/sonarqube-server/9.9/analyzing-source-code/languages/javascript-typescript-css},
  urldate = {2025-12-05},
  langid = {english}
}

@misc{CodeQLAPIGraph,
  author = {GitHub},
  title = {{API} Graph Documentation for {JavaScript}},
  year = {2025},
  howpublished = {\url{https://codeql.github.com/codeql-standard-libraries/javascript/semmle/javascript/ApiGraphs.qll/module.ApiGraphs.html}},
  note = {CodeQL Standard Libraries}
}

@misc{CODEQL_DEBUGGING,
  author = {GitHub},
  title = {Debugging Data Flow Queries Using Partial Flow},
  year = {2025},
  howpublished = {\url{https://codeql.github.com/docs/writing-codeql-queries/debugging-data-flow-queries-using-partial-flow/}},
  note = {CodeQL Docs}
}

@misc{CodeQLPropRules,
  author = {GitHub},
  title = {Analyzing data flow in JavaScript and TypeScript},
  year = {2025},
  howpublished = {\url{https://codeql.github.com/docs/codeql-language-guides/analyzing-data-flow-in-javascript-and-typescript}},
  note = {CodeQL Docs}
}

@misc{CodeQLCWECoverage,
  author = {GitHub},
  title = {{CWE} Coverage for {JavaScript} and {TypeScript}},
  year = {2025},
  howpublished = {\url{https://codeql.github.com/codeql-query-help/javascript-cwe/}},
  note = {CodeQL Query Help Documentation}
}

@misc{swebench-verified,
  title = {{SWE-bench Verified}},
  author = {{Epoch AI}},
  year = {2024},
  howpublished = {\url{https://epoch.ai/benchmarks/swe-bench-verified}},
  note = {Evaluation methodology enforces a 1 million token limit per issue}
}

@misc{LANGCHAIN,
  title={LangChain: Build context-aware reasoning applications},
  author={LangChain, Inc.},
  note={\url{https://www.langchain.com/}},
  year={2024}
}

@online{GHSA-5ff8-jcf9-fw62,
  title = {{{GHSA-5ff8-jcf9-fw62}} - {{GitHub Advisory Database}}},
  url = {https://github.com/advisories/GHSA-5ff8-jcf9-fw62},
  urldate = {2025-12-13},
  langid = {english},
  organization = {GitHub}
}

@online{GHSA-hpr5-wp7c-hh5q,
  title = {{{GHSA-hpr5-wp7c-hh5q}} - {{GitHub Advisory Database}}},
  url = {https://github.com/advisories/GHSA-hpr5-wp7c-hh5q},
  urldate = {2025-12-13},
  langid = {english},
  organization = {GitHub}
}

@online{RE-ACT,
  title = {{{ReAct}}: {{Synergizing Reasoning}} and {{Acting}} in {{Language Models}}},
  shorttitle = {{{ReAct}}},
  author = {Yao, Shunyu and Zhao, Jeffrey and Yu, Dian and Du, Nan and Shafran, Izhak and Narasimhan, Karthik and Cao, Yuan},
  date = {2023-03-10},
  eprint = {2210.03629},
  eprinttype = {arXiv},
  eprintclass = {cs},
  doi = {10.48550/arXiv.2210.03629},
  url = {http://arxiv.org/abs/2210.03629},
  urldate = {2025-12-15},
  pubstate = {prepublished}
}

@misc{CODEQL_REPOS,
  title   = {CodeQL updates from the first half of 2023 - GitHub Changelog},
  note    = {\url{https://github.blog/changelog/2023-07-26-codeql-updates-from-the-first-half-of-2023/}},
  journal = {GitHub Changelog},
  author  = {GitHub},
  year    = {2023}
}

@misc{LLM_COT,
      title={Chain-of-Thought Prompting Elicits Reasoning in Large Language Models}, 
      author={Jason Wei and Xuezhi Wang and Dale Schuurmans and Maarten Bosma and Brian Ichter and Fei Xia and Ed Chi and Quoc Le and Denny Zhou},
      year={2023},
      eprint={2201.11903},
      archivePrefix={arXiv},
      primaryClass={cs.CL},
      howpublished={https://arxiv.org/abs/2201.11903}, 
}

@misc{Semgrep_Main_Docs,
  title   = {Semgrep App Security Platform | AI-assisted SAST, SCA and Secrets Detection},
  note    = {\url{https://semgrep.dev/}},
  journal = {Semgrep},
  author  = {Semgrep},
  year    = {2025}
}

@misc{SonarQube_Main_Docs,
  title   = {SonarQube | Code Quality \& Security | Static Analysis Tool | Sonar},
  note    = {\url{https://www.sonarsource.com/products/sonarqube/}},
  journal = {SonarSource},
  author  = {SonarSource},
  year    = {2025}
}

@article{SENECA,
   title={Seneca: Taint-Based Call Graph Construction for Java Object Deserialization},
   volume={8},
   ISSN={2475-1421},
   url={http://dx.doi.org/10.1145/3649851},
   DOI={10.1145/3649851},
   number={OOPSLA1},
   journal={Proceedings of the ACM on Programming Languages},
   publisher={Association for Computing Machinery (ACM)},
   author={Santos, Joanna C. S. and Mirakhorli, Mehdi and Shokri, Ali},
   year={2024},
   month=apr, pages={1125–1153} }

@article{TAJ,
author = {Tripp, Omer and Pistoia, Marco and Fink, Stephen J. and Sridharan, Manu and Weisman, Omri},
title = {TAJ: effective taint analysis of web applications},
year = {2009},
issue_date = {June 2009},
publisher = {Association for Computing Machinery},
address = {New York, NY, USA},
volume = {44},
number = {6},
issn = {0362-1340},
url = {https://doi.org/10.1145/1543135.1542486},
doi = {10.1145/1543135.1542486},
journal = {SIGPLAN Not.},
month = jun,
pages = {87–97},
numpages = {11}
}

@misc{AGENT_ROLES1,
      title={ChatDev: Communicative Agents for Software Development}, 
      author={Chen Qian and Wei Liu and Hongzhang Liu and Nuo Chen and Yufan Dang and Jiahao Li and Cheng Yang and Weize Chen and Yusheng Su and Xin Cong and Juyuan Xu and Dahai Li and Zhiyuan Liu and Maosong Sun},
      year={2024},
      eprint={2307.07924},
      archivePrefix={arXiv},
      primaryClass={cs.SE},
      howpublished={https://arxiv.org/abs/2307.07924}, 
}

\appendices
\pagebreak
\section{Agent Walkthrough}
\label{sec:agent-walkthrough}

\subsection{Agent Implementation}
\label{sec:agent-implementation}
\toolname's agents are implemented using LangChain~\cite{LANGCHAIN}, a framework that provides a unified interface across LLM providers. Each agent follows a common architecture: prompts are constructed as system/user message pairs from Jinja2 templates, domain-specific tools are defined with Pydantic schemas for input validation, and the LLM is invoked in an iterative loop where tool calls are executed and results returned until the agent signals completion or reaches its iteration limit. All agents share a read-only codebase traversal toolkit for exploration, while each adds task-specific resolution tools.

\myparagraph{Tool Calls}

The \textit{traversal toolkit} is available to all agents as a read-only set of tools for codebase exploration. These \sourceSinkAgentName and \flowSummaryAgentName tools operate on multiple codebases under namespace prefixes (e.g., \texttt{source/}, \texttt{npm/}, \texttt{builtin/}), enabling unified navigation across package source code, npm dependencies, and Node.js built-in modules. In contrast, the \callgraphAgentName Agent tools only require the \texttt{source/} namespace:

\begin{itemize}[leftmargin=*, nosep]
    \item \texttt{view\_src(file\_path, start\_line, end\_line)}: Retrieves source code lines with line numbers from indexed files.
    \item \texttt{view\_dir(path)}: Returns a tree representation of the project structure, optionally scoped to a subdirectory.
    \item \texttt{find\_string(search, max\_results, start\_index)}: Searches for string occurrences across all indexed files with pagination 
    support.
\end{itemize}

The \textit{\sourceSinkAgentName Agent} contains custom tools for proposing taint entry points and dangerous operations:

\begin{itemize}[leftmargin=*, nosep]
    \item \texttt{propose\_source(location, description, confidence)}: Proposes a code location where untrusted input enters the application. 
    \item \texttt{propose\_sink(location, description, confidence)}: Proposes a code location where a dangerous operation occurs.
    \item \texttt{view\_proposed\_sources\_sinks(verbose, item\_type)}: Reviews already-proposed sources and sinks to avoid duplicates and assess coverage.
    \item \texttt{complete\_discovery(justification, summary, confidence)}: Signals discovery completion with an explanation of search strategies used.
\end{itemize}

The \textit{\callgraphAgentName Agent} contains custom tools for resolving unresolved call sites to their target functions:

\begin{itemize}[leftmargin=*, nosep]
    \item \texttt{search\_functions(function\_name, file\_path)}: Searches the CodeQL-extracted function index by name or file path with fuzzy matching, returning function indices required for proposals.
    \item \texttt{propose\_fp(candidates)}: Proposes one or more first-party callees with a traced path from call site to target, including intermediate steps through imports, assignments, and lookups.
    \item \texttt{propose\_tp(library\_name, metadata, confidence, reasoning)}: Marks the call as targeting a third-party library with structured metadata including library type, module path, and import statement.
    \item \texttt{mark\_target\_not\_indexed(target\_file, target\_line, target\_name, explanation)}: Records targets found in source code but missing from the function index due to CodeQL extraction limitations.
    \item \texttt{view\_proposed\_callees(verbose)}: Reviews already-proposed callees to avoid duplicates.
    \item \texttt{complete\_resolution(status, summary)}: Signals resolution completion as either \texttt{resolved} or \texttt{unresolvable}.
\end{itemize}

The \textit{\flowSummaryAgentName Agent} contains custom tools for validating third-party edges in vulnerability paths:

\begin{itemize}[leftmargin=*, nosep]
    \item \texttt{classify\_edge(flow\_trace)}: Classifies a third-party edge as \texttt{propagates-taint}, \texttt{sanitizes-taint}, or \texttt{unknown} with a complete data flow trace through the library code, including sanitization points and confidence score.
\end{itemize}

\myparagraph{Multi-Run Aggregation}
Due to LLM non-determinism, each agent executes multiple runs per task, with results combined through agent-specific aggregation strategies. 

\begin{itemize}
    \item The \sourceSinkAgentName Agent uses union aggregation: sources and sinks are grouped by location identity (\texttt{file:line:string\_value}), and any fact observed in at least one run is retained, maximizing recall.
    \item The \callgraphAgentName Agent uses stratified aggregation: a majority vote first determines whether each call site targets first-party or third-party code, then first-party edges undergo union aggregation since the LLM may resolve a call to different targets across runs, while third-party edges require no further aggregation as candidate flow summaries are deterministic per call site.
    \item  The \flowSummaryAgentName Agent uses majority voting across an odd number of runs to classify edges as \texttt{propagates}, \texttt{sanitizes}, or \texttt{unknown}, with ties broken by average confidence scores; unknown classifications are conservatively treated as propagating taint.
\end{itemize}




\subsubsection{Confidence}
\label{appendix:confidence}
To mitigate LLM overconfidence, each agent employs calibrated rubrics.
All rubrics reserve score~5 for zero-ambiguity cases, require justification against documented evidence, and instruct agents to prefer lower scores when uncertain.

\textbf{CallGraph Agent:} Scores (1--5) reflect trace completeness: 5 requires direct invocation with unambiguous resolution; 4 permits dynamic behavior if multiple evidence points converge; 3--1 indicate increasing gaps or assumptions in the trace.

\textbf{Source/Sink Agent:} For sources, scores reflect user-controllability and absence of entry-point sanitization. For sinks, scores reflect known exploitability and whether exploit preconditions are evidenced. Both reserve 5 for certain, unmitigated vulnerability paths.

\textbf{Flow Summary Agent:} Scores reflect evidence strength for propagation classification: 5 requires direct implementation inspection or authoritative documentation; lower scores indicate increasing reliance on semantic inference.

\begin{figure*}[t]
    \centering
        \includegraphics[
        width=\linewidth,
        trim=0 0 0 0,  
    ]{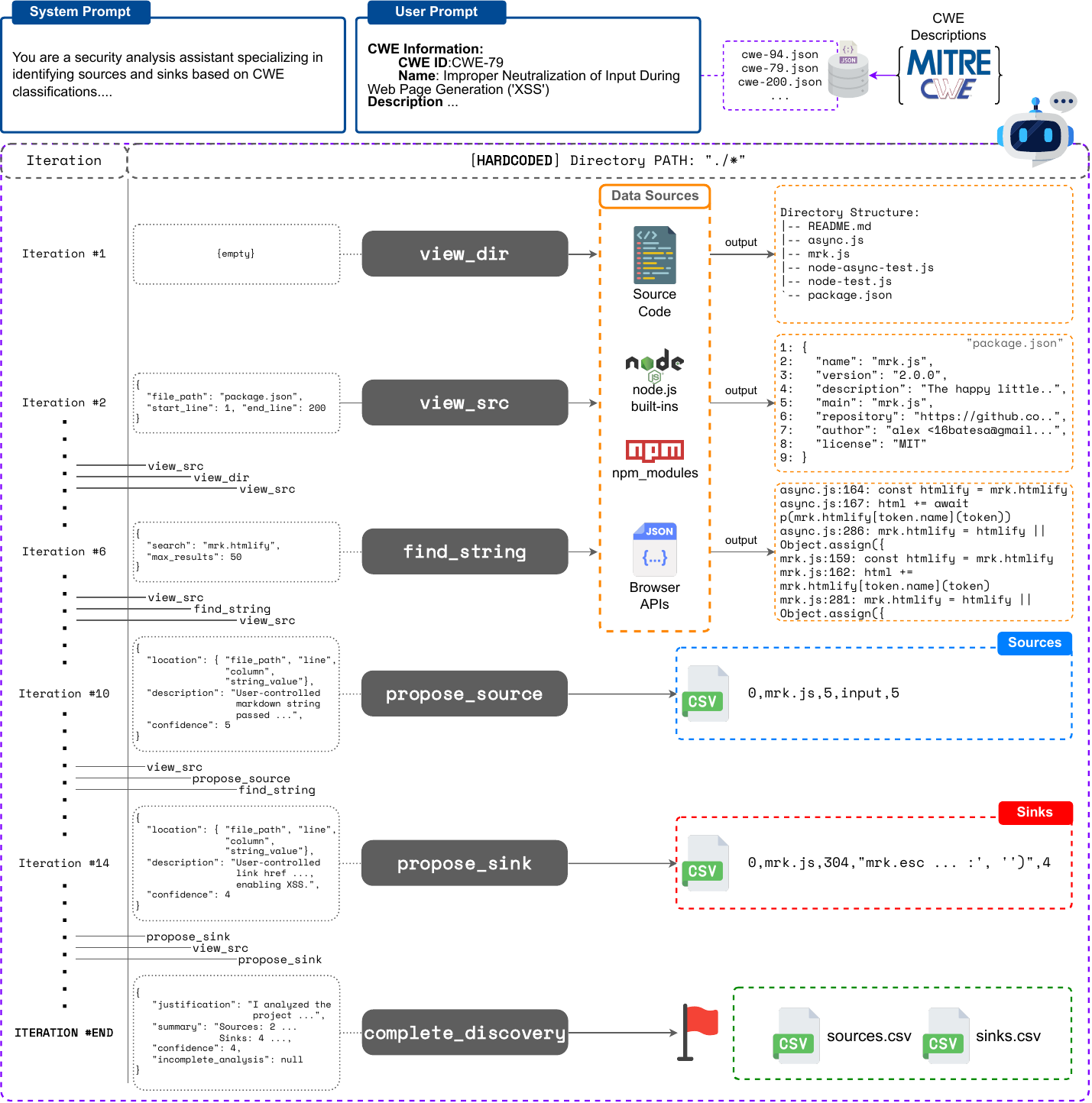}%
    \centering
    \caption{\sourceSinkAgentName Agent discovering XSS sources and sinks in mrk.js through iterative code exploration.}
    \label{fig:source_sink_agent}
\end{figure*}

\subsection{\sourceSinkAgentName Agent Walkthrough}
\label{sec:source-sink-agent-walkthrough}
Figure~\ref{fig:source_sink_agent} illustrates a single run of the Source/Sink Agent's analysis of \texttt{mrk.js} for CWE-79 (Cross-site Scripting). The agent receives a structured CWE pattern report containing the vulnerability description, common consequences, and characteristic patterns for XSS.

The agent begins by exploring the project structure using \texttt{view\_dir}, revealing a minimal package: two implementation files (\texttt{mrk.js}, \texttt{async.js}), test files, and configuration. It then examines \texttt{package.json} to understand that this is a markdown parser that renders user input to HTML, a context where XSS vulnerabilities commonly arise.

With this understanding, the agent systematically reviews the implementation. Using \texttt{find\_string} to search for \texttt{mrk.htmlify}, it locates the HTML rendering functions where user-controlled markdown tokens are converted to HTML output. The agent traces data flow from the entry point backward: the \texttt{mrk(input)} function parameter at line~5 represents where untrusted markdown strings enter the parser.

For sinks, the agent identifies the \texttt{htmlify.link} function at line~304, which interpolates user-controlled \texttt{metadata.href} into an HTML \texttt{href} attribute. The agent recognizes that the sanitization, \texttt{.replace('javascript:', '')}, is case-sensitive and bypassable, constituting an XSS sink with confidence~4.

After reviewing its proposals with \texttt{view\_proposed\_sources\_sinks} to avoid duplicates, the agent calls \texttt{complete\_discovery}, reporting 2~sources (the \texttt{input} parameters in both sync and async variants) and 4~sinks (link and autolink rendering in both files) with 4 confidence. 
These location tuples are then validated against CodeQL's QL~Nodes to produce the final source and sink specifications shown in Figure~\ref{fig:TaintSpecs}.

\begin{figure*}[t]
    \centering
        \includegraphics[
        width=\linewidth,
        trim=0 0 0 0,  
    ]{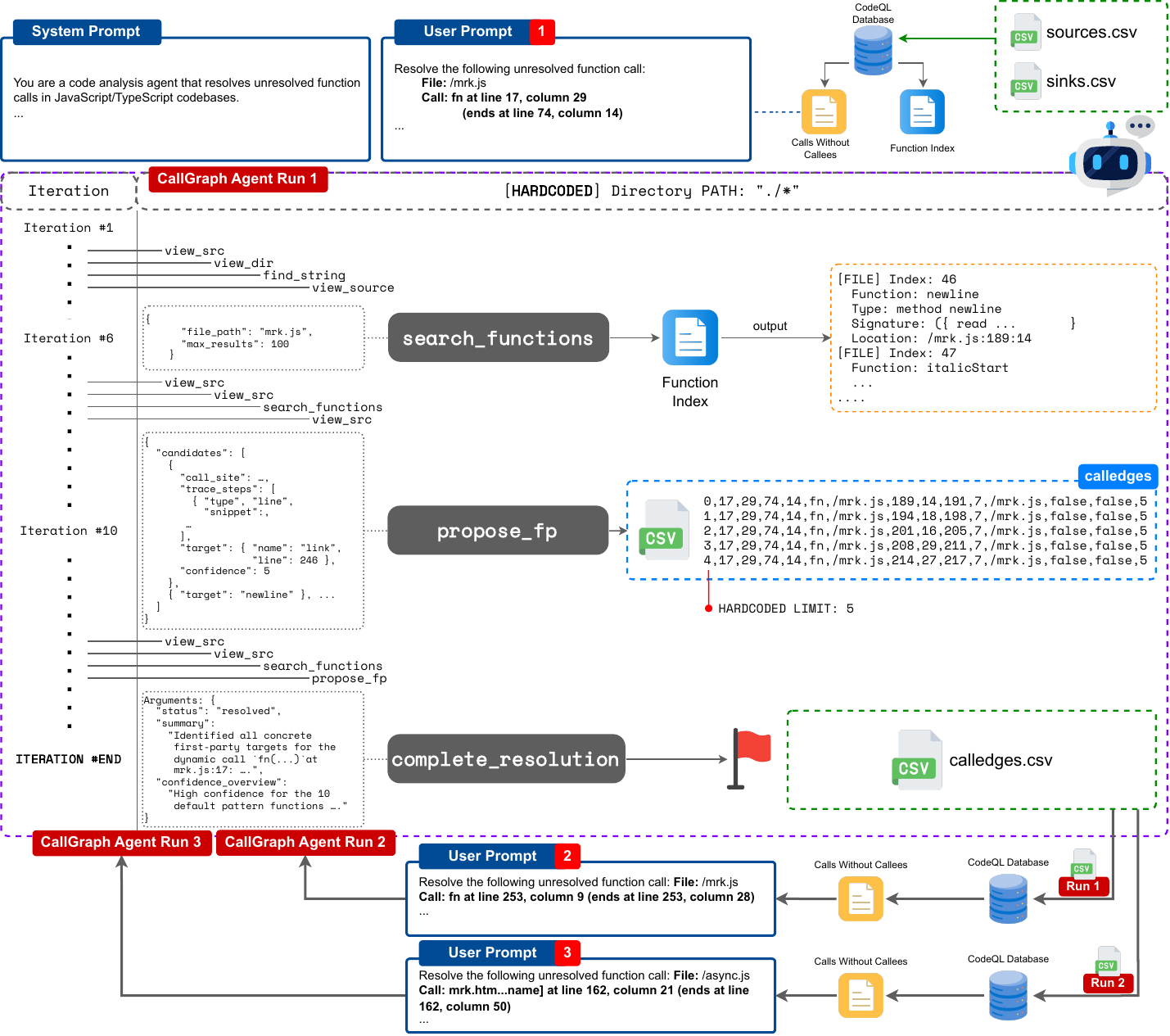}%
    \centering
    \caption{\callgraphAgentName Agent resolution of unresolved calls in mrk.js, illustrating iterative code exploration of a dynamic dispatch call in Listing~\ref{code:mrk_strat_pattern}}
    \label{fig:CallGraph_Agent}
\end{figure*}

\subsection{\callgraphAgentName Agent Walkthrough}
\label{appendix:callgraph-agent}
Figure~\ref{fig:CallGraph_Agent} illustrates the \callgraphAgentName Agent resolving the three unresolved calls identified in Section~\ref{sec:motivation}. The agent receives each unresolved call as a separate task: the file path, line and column location, callee name (when available), and access to the function index extracted from CodeQL.

We trace one resolution in detail: the dynamic dispatch at line~17 (Listing~\ref{code:mrk_strat_pattern}), where \texttt{fn} is invoked in a for loop. CodeQL cannot resolve this call because \texttt{fn} is bound dynamically from object property enumeration.

The agent begins by calling \texttt{view\_src} to examine the surrounding code, revealing that \texttt{fn} iterates over \texttt{mrk.patterns}. Using \texttt{view\_src}, it traces where \texttt{mrk.patterns} is assigned via \texttt{Object.assign} with default pattern functions. The agent then calls \texttt{search\_functions} with \texttt{file\_path="mrk.js"} to retrieve all function indices in the file.

For each pattern function (\texttt{newline}, \texttt{italicStart}, \texttt{link}, etc.), the agent constructs a \texttt{TracedCandidateInfo} structure documenting: (1)~the call site location, (2)~trace steps showing assignment through \texttt{Object.entries} iteration, property lookup in \texttt{mrk.patterns}, and the target function definition, (3)~the function index required for CodeQL binding, and (4)~a confidence score. 
The agent proposes these 10 pattern functions via \texttt{propose\_fp} in batches of at most 5; larger batches degraded trace quality as the agent produced shallower reasoning and structural output errors. Each candidate receives confidence~5 due to the unambiguous trace from call site to definition.

\begin{figure}[t]
    \centering
    \includegraphics[width=\linewidth]{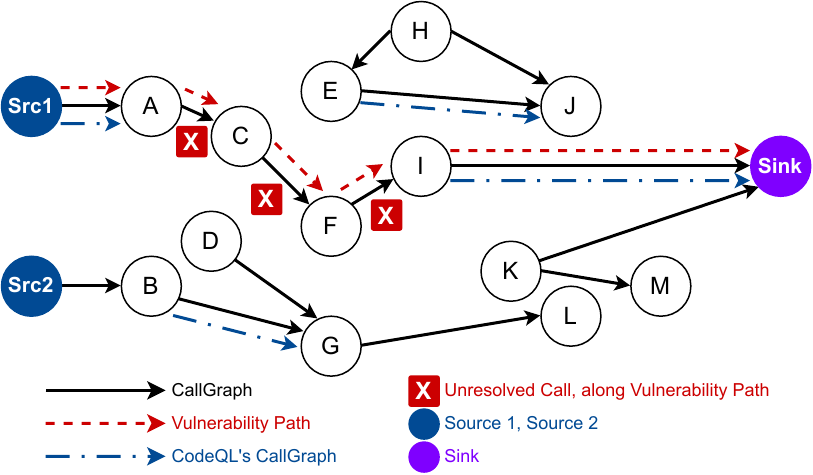}
    \caption{Taint-flow graph illustrating TICR's demand-driven approach}
    \label{fig:taint-vuln-path}
\end{figure}

\myparagraph{\callgraphAgentName with TICR} Figure~\ref{fig:taint-vuln-path} illustrates how TICR leverages resolved edges to expose additional security-relevant calls. In the initial iteration, TICR identifies calls reachable from sources (the \texttt{input} parameter) but cannot reach sinks because dynamic dispatch severs the path. After the CallGraph Agent resolves \texttt{fn}$\rightarrow$\texttt{link}, CodeQL can propagate taint through the pattern functions. This exposes the callback invocation \texttt{meta(\{name,~href\})} at line~253 (``Break~2''), which TICR adds to the next iteration's worklist. Resolving this edge in turn enables taint to reach the \texttt{htmlify} dispatch (``Break~3''), ultimately connecting the source to the XSS sink. This iterative refinement continues until no new security-relevant calls emerge or the iteration limit is reached.

\begin{figure}[t]
\centering
    \includegraphics[
    width=\linewidth,
    trim=0 0 0 0,  
]{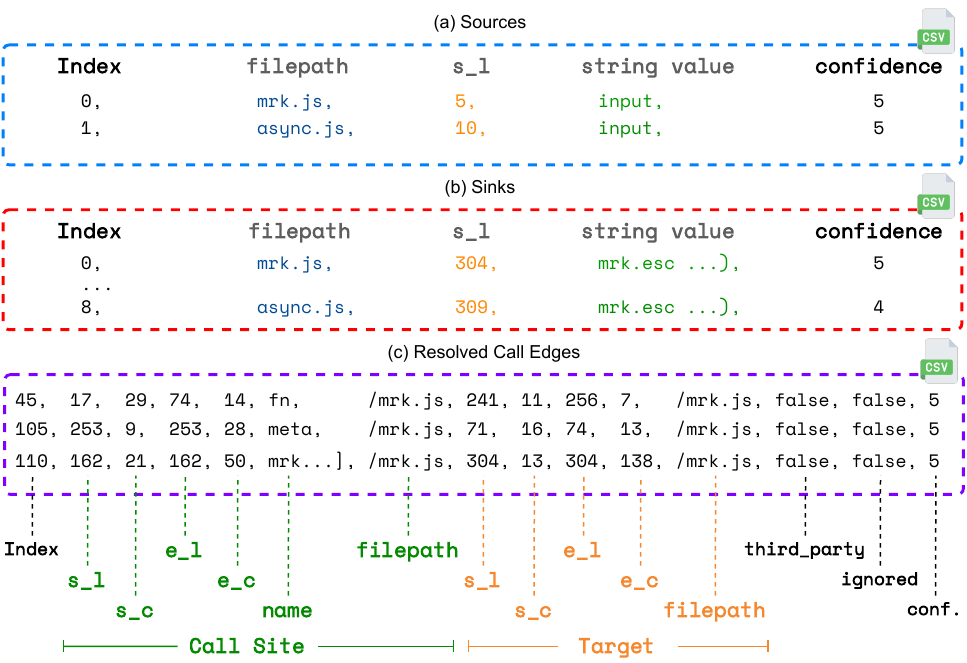}%
    \caption{Extracted Taint Specifications for \texttt{mrk.js}}
    \label{fig:TaintSpecs}
\end{figure}

The final \texttt{calledges.csv}, shown in Figure~\ref{fig:TaintSpecs}, contains the recovered edges that repair CodeQL's inter-procedural data flow graph.

\begin{figure}[t]
\centering
    \includegraphics[
    width=\linewidth,
    trim=0 0 0 0,  
]{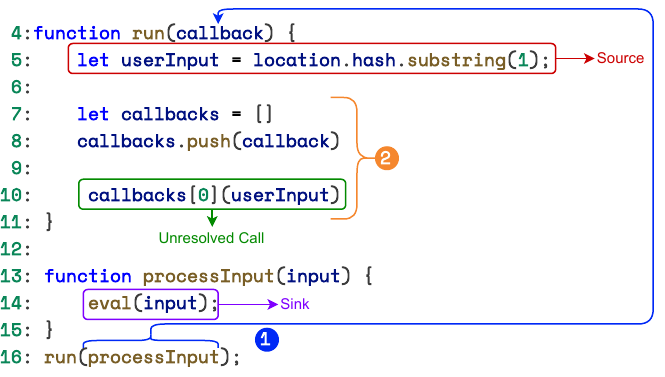}%
    \caption{Source-to-Break: Taint flows \textit{into} the broken call.}
    \label{fig:Source-Break}
\end{figure}

\begin{figure}[t]
\centering
    \includegraphics[
    width=\linewidth,
    trim=0 0 0 0,  
]{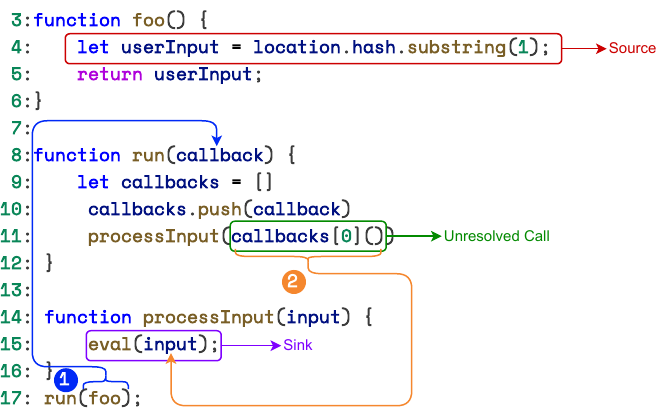}%
    \caption{Break-to-Sink: Taint flows \textit{out of} the broken call.}
    \label{fig:Break-Sink}
\end{figure}

\subsection{Flow Summary Agent Walkthrough}
\label{appendix:flow-summary-agent-walkthrough}

The mrk.js example from Section~\ref{sec:motivation} did not contain third-party library calls along its vulnerability path, allowing the CallGraph agent to resolve all inter-procedural edges within the package's source code.
However, many real-world vulnerabilities traverse npm dependencies, where taint flows through third-party APIs before reaching dangerous sinks.
To illustrate how \toolname handles these cases, we examine GHSA-5ff8-jcf9-fw62~\cite{GHSA-5ff8-jcf9-fw62}, a cross-site scripting vulnerability in \texttt{markdown-it-katex}~\cite{markdown-it-katex}, a plugin that adds mathematical formula rendering to the popular \texttt{markdown-it} parser.

\begin{figure*}[t]
\centering
    \includegraphics[
    width=\linewidth,
    trim=0 0 0 0,  
]{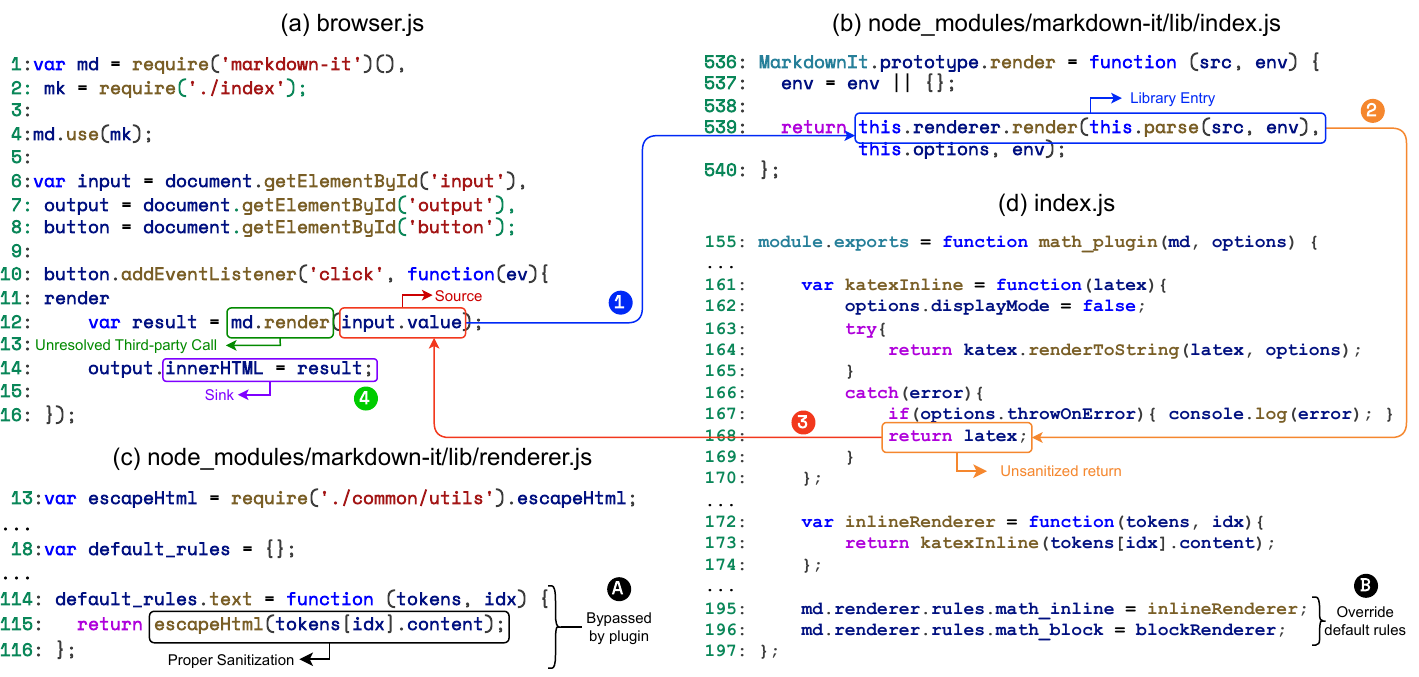}%
    \caption{Cross-file vulnerability path in markdown-it-katex, requiring analysis of third-party APIs}
    \label{fig:Flow_summary_Example}
\end{figure*}

\myparagraph{Vulnerability Context}
Figure~\ref{fig:Flow_summary_Example} illustrates the vulnerability across four code segments: the application entry point (\texttt{browser.js}), the vulnerable plugin (\texttt{index.js}), and two files from the \texttt{markdown-it} dependency in \texttt{node\_modules}.
The application reads user input from a text field and passes it to \texttt{md.render(input.value)} \circletext{custombluedark}{1}, where both the taint source and the unresolved third-party call reside.
The rendered result flows to \texttt{innerHTML} \circletext{customgreen}{4}, an XSS sink.

CodeQL identifies both the source (\texttt{input.value}) and sink (\texttt{innerHTML}) but cannot complete the vulnerability path.
The call to \texttt{md.render()} crosses into \texttt{node\_modules/markdown-it}, which CodeQL excludes from analysis by default.
Without a manually specified flow summary for this API, CodeQL conservatively assumes the call blocks taint propagation, yielding no vulnerability alert.

\myparagraph{Candidate Flow Summary}
When the CallGraph agent encounters \texttt{md.render()}, it recognizes the call targets a third-party library (via CodeQL's API graph) and marks it as a \emph{candidate flow summary}.
This enables CodeQL to report a candidate vulnerability path, which \toolname then validates through the \flowSummaryAgentName Agent.

\begin{figure*}[t]
\centering
    \includegraphics[
    width=\linewidth,
    trim=0 0 0 0,  
]{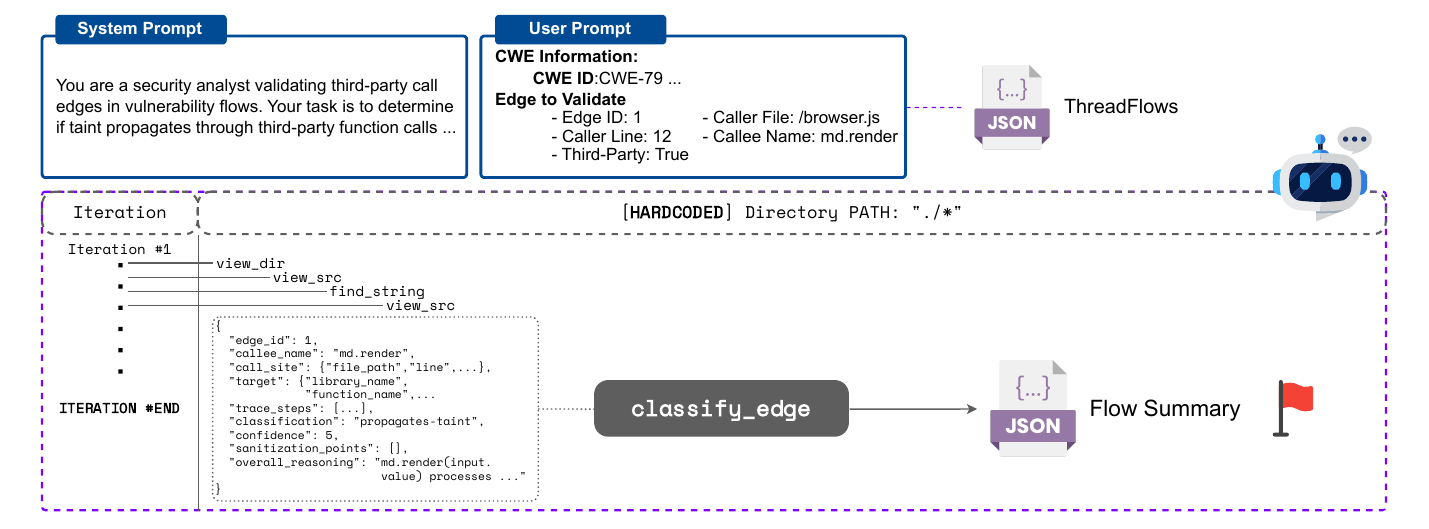}%
    \caption{Flow Summary Agent analysis classifying the third-party call as taint-propagating}
    \label{fig:Flow_Summary_Agent}
\end{figure*}

\myparagraph{Agent Analysis}
The \flowSummaryAgentName Agent receives the candidate edge along with CWE-79 context and begins systematic exploration.
Figure~\ref{fig:Flow_Summary_Agent} shows a high-level overview of the agent's tool calls and reasoning.

The agent first examines the call site in \texttt{browser.js}, identifying that \texttt{md} is instantiated from \texttt{markdown-it} and extended via \texttt{md.use(mk)} with a local plugin.
Using the \texttt{npm/} namespace prefix, the agent navigates into \texttt{node\_modules} to examine the library's implementation.

At \texttt{npm/markdown-it/lib/index.js} \circletext{customorange}{2}, the agent locates the \texttt{render} function (line~536--540), which parses input into tokens and delegates to \texttt{this.renderer.render()}.
The agent then searches for sanitization mechanisms, finding \texttt{escapeHtml} usage throughout the renderer.
At \circletext{black}{A} in \texttt{renderer.js}, the agent observes that markdown-it's default text rule applies HTML entity encoding.
This would neutralize XSS payloads if it were applied.
However, the agent also examines the local plugin (\texttt{source/index.js}), discovering that it \textit{\emph{overrides}} markdown-it's renderer rules at \circletext{black}{B}.

These custom rules bypass the default sanitization entirely.
The agent traces into \texttt{katexInline} \circletext{customred}{3} and identifies the critical vulnerability: when KaTeX fails to render a LaTeX expression, the error handler returns the raw input without escaping. 

\myparagraph{Classification}
The agent classifies the edge as \texttt{propagates-taint} with confidence~5, documenting seven trace steps from call site through library entry, plugin registration, renderer override, the unsanitized error path, and library exit.
Because the candidate flow summary propagates taint, the vulnerability alert remains valid in \toolname's final results.

\section{TICR Implementation}
\label{appendix:ticr-implementation}

Section~\ref{sec:ticr} introduced TICR's bidirectional analysis and defined the reachability sets $M_{src \to brk}$ and $M_{brk \to snk}$. This appendix explains why both directions are necessary and illustrates the implementation.

\subsection{Bidirectional Analysis}

A unidirectional approach, tracking only from sources toward broken calls, or only from broken calls toward sinks, misses security-relevant calls depending on where taint enters or exits the unresolved invocation.

\mysubtitle{Pattern 1: Taint Flows INTO the Broken Call}
Figure~\ref{fig:Source-Break} illustrates a pattern where taint flows directly into an unresolved call. The function \texttt{processInput}, which contains the \texttt{eval} sink, is passed as a callback argument to \texttt{run()} \circletext{custombluedark}{1}. Inside \texttt{run()}, this callback is stored in an array and later invoked via \texttt{callbacks[0](userInput)} \circletext{customorange}{2}, an indirect call that CodeQL cannot statically resolve. Tainted data from the source \texttt{location.hash} flows through \texttt{userInput} directly into this invocation's argument.

Source-to-break analysis  $\mathcal{M}_{\text{src→brk}}$~(Equation~\ref{eq:ticr_src_to_brk}) detects this pattern because taint propagates from the source \emph{to} the unresolved call site. Break-to-sink analysis misses it: the sink resides \emph{inside} the unresolved callee \texttt{processInput}.

\mysubtitle{Pattern 2: Taint Flows OUT OF the Broken Call}
Figure~\ref{fig:Break-Sink} illustrates the complementary pattern where taint flows out of a broken call. The function \texttt{foo}, which reads from the source \texttt{location.hash} and returns tainted data, is passed as a callback to \texttt{run()} \circletext{custombluedark}{1}. Inside \texttt{run()}, this callback is stored in an array and invoked via \texttt{callbacks[0]()}, an indirect call that CodeQL cannot resolve. The return value of this unresolved call flows directly into \texttt{processInput()}, reaching the \texttt{eval} sink \circletext{customorange}{2}.

Break-to-sink analysis $\mathcal{M}_{\text{brk→snk}}$~(Equation~\ref{eq:ticr_brk_to_snk}) detects this pattern because taint propagates \emph{from} the unresolved call site to the sink. Source-to-break analysis misses it: no tainted data flows into the call's arguments.

The union strategy $\mathcal{M}_{\text{TICR}} = \mathcal{M}_{\text{src→brk}} \cup \mathcal{M}_{\text{brk→snk}}$~(Equation~\ref{eq:ticr_union}) ensures both patterns are captured, identifying all unresolved calls that could plausibly bridge a source-to-sink gap.









\section{Additional Figures and Tables}
\label{sec:additonal-figures}

\definecolor{sectioncolor}{RGB}{0, 51, 102}
\definecolor{subsectioncolor}{RGB}{70, 70, 70}
\definecolor{instructionbg}{RGB}{245, 245, 245}
\definecolor{instructionframe}{RGB}{180, 180, 180}

\begin{table}[H]
\centering
\footnotesize
\caption{Validation Reference and Model Comparison Results}
\label{table:val-model-combined}

\resizebox{\columnwidth}{!}{%
\begin{tabular}{
@{\hspace{5pt}}c
@{\hspace{10pt}}l
@{\hspace{20pt}}l
@{\hspace{10pt}}c
@{\hspace{10pt}}c@{\hspace{5pt}}
}
\toprule
\textbf{ID} & \textbf{GHSA} & \textbf{CWE} & \textbf{Gemini} & \textbf{GPT} \\
\midrule
1  & GHSA-3crj-w4f5-gwh4 & CWE-74  & x          & \checkmark \\
2  & GHSA-5ff8-jcf9-fw62 & CWE-79  & \checkmark & \checkmark \\
3  & GHSA-8j8c-7jfh-h6hx & CWE-94  & x          & \checkmark \\
4  & GHSA-cqjg-whmm-8gv6 & CWE-400 & x          & x \\
5  & GHSA-f4hq-453j-p95f & CWE-601 & \checkmark & \checkmark \\
6  & GHSA-hpr5-wp7c-hh5q & CWE-79  & x          & \checkmark \\
7  & GHSA-jv35-xqg7-f92r & CWE-915 & \checkmark & \checkmark \\
8  & GHSA-r4m5-47cq-6qg8 & CWE-918 & \checkmark & \checkmark \\
9  & GHSA-x67x-98x7-wv26 & CWE-78  & \checkmark & \checkmark \\
10 & GHSA-xqh8-5j36-4556 & CWE-89  & \checkmark & \checkmark \\
\midrule
\multicolumn{3}{l}{\textbf{Recall}} & 60\% & \textbf{90\%} \\
\multicolumn{3}{l}{\textbf{Cost}}   & \textbf{\$77} & \$92.75 \\
\bottomrule
\end{tabular}
}

\normalsize
\end{table}


\end{document}